\documentclass[twocolumn]{aastex62} 
\pdfoutput=1
\usepackage{amsmath}

\usepackage{hyperref} \usepackage{color}

\newcommand{\kpc}{\,{\rm kpc}} 
\newcommand{\am}[0]{[$\alpha$/{\rm M]}} 
\newcommand{\mh}[0]{[{\rm M/H}]}
\newcommand{\xh}[1]{[{\rm #1/H}]}

\newcommand{\xfe}[1]{[{\rm #1/Fe}]}
\newcommand{\xmg}[1]{[{\rm #1/Mg}]} 
\newcommand{\afe}[0]{[$\alpha$/{\rm Fe}]}
\newcommand{\feh}[0]{[{\rm Fe/H}]} 
\newcommand{\mgfe}[0]{[{\rm Mg/Fe}]} 
\newcommand{\femg}{[{\rm Fe}/{\rm Mg}]} 
\newcommand{\mgh}{[{\rm Mg}/{\rm H}]}

\newcommand{\teff}[0]{$T_{\rm eff}$}
\newcommand{\logg}[0]{$\log{g}$}

\newcommand{\Acc}{A_{\rm cc}}
\newcommand{\AIa}{A_{\rm Ia}} 
\newcommand{\pxcc}{p_{\rm cc}^X}
\newcommand{\pxIa}{p_{\rm Ia}^X} 
\newcommand{\pxccs}{p_{\rm cc,\odot}^X}
\newcommand{\pxIas}{p_{\rm Ia,\odot}^X}

\newcommand{\alphacc}{\alpha_{\rm cc}}
\newcommand{\alphaIa}{\alpha_{\rm Ia}}
\newcommand{\Rxsun}{R^X_{\rm Ia}} 
\newcommand{\Zsun}{Z_\odot}
\newcommand{\mdotstar}{\dot{M}_*}
\newcommand{\tausfh}{\tau_{\rm sfh}}
\newcommand{\taustar}{\tau_{*}}
\newcommand\Gyr{\,{\rm Gyr}}
\newcommand{\K}{\,{\rm K}}

\begin{document}

\title{Chemical Cartography with APOGEE: Multi-element abundance ratios}

 \author[0000-0001-7775-7261]{David H. Weinberg}
 \affiliation{Department of Astronomy and CCAPP, The Ohio State University,
 Columbus, OH 43210, USA}

 \author[0000-0002-9771-9622]{Jon A. Holtzman}
\affiliation{New Mexico State University, Las Cruces, NM 88003, USA}

\author{Sten Hasselquist}
\affiliation{New Mexico State University, Las Cruces, NM 88003, USA}

\author{Jonathan C.  Bird}
\affiliation{Vanderbilt University, Nashville, TN 37240, USA}

\author{Jennifer A. Johnson}
\affiliation{Department of Astronomy and CCAPP, The Ohio State University, Columbus, OH 43210, USA}

\author[0000-0003-0509-2656]{Matthew Shetrone }
\affiliation{University of Texas at Austin, McDonald Observatory, Fort Davis, TX 79734, USA}

\author{Jennifer Sobeck}
\affiliation{Department of Astronomy, Box 351580, University of Washington, Seattle, WA 98195, USA}

\author{Carlos Allende Prieto}
\affiliation{Instituto de Astrof\'isica de Canarias (IAC), E-38205 La Laguna, Tenerife, Spain}
\affiliation{Universidad de La Laguna, Dpto. Astrof\'isica, E-38206 La Laguna, Tenerife, Spain}

\author{Dmitry Bizyaev}
\affiliation{Apache Point Observatory, P.O. Box 59, Sunspot, NM 88349, USA}

\author{Ricardo Carrera}
\affiliation{Astronomical Observatory of Padova, National Institute of 
Astrophysics, Vicolo Osservatorio 5 - 35122 - Padova, Italy}

\author{Roger E.\ Cohen}
\affiliation{Departamento de Astronom\'ia, Universidad de Concepci\'on, 
Casilla 160-C, Concepci\'on, Chile}
\affiliation{Space Telescope Science Institute, 3700 San Martin Drive, Baltimore, MD 21210, USA}

\author{Katia Cunha}
\affiliation{Observat\'orio Nacional, S\~ao Crist\'ov\~ao, Rio de Janeiro, Brazil}
\affiliation{University of Arizona, Tucson, AZ 85719, USA}

\author{Garrett Ebelke}
\affiliation{Department of Astronomy, University of Virginia, Charlottesville,
VA 22904, USA}

\author{J.\ G.\ Fern\'andez-Trincado}
\affiliation{Instituto de Astronom\'ia y Ciencias Planetarias, 
Universidad de Atacama, Copayapu 485, Copiap\'o, Chile}
\affiliation{Institut Utinam, CNRS UMR 6213, Universit\'e 
Bourgogne-Franche-Comt\'e, OSU THETA Franche-Comt\'e, 
Observatoire de Besan\c{c}on, \\ BP 1615, 25010 Besan\c{c}on Cedex, France}
\affiliation{Departamento de Astronom\'\i a, Casilla 160-C, 
Universidad de Concepci\'on, Concepci\'on, Chile}

\author{D.\ A.\ Garc\'ia-Hern\'andez}
\affiliation{Instituto de Astrof\'isica de Canarias (IAC), E-38205 La Laguna, Tenerife, Spain}
\affiliation{Universidad de La Laguna, Dpto. Astrof\'isica, E-38206 La Laguna, Tenerife, Spain}

\author{Christian R.\ Hayes}
\affiliation{Department of Astronomy, University of Virginia, Charlottesville,
VA 22904, USA}

\author[0000-0002-4912-8609]{Henrik J\"onsson}
\affiliation{Lund Observatory, Department of Astronomy and Theoretical Physics, Lund University, Box 43, SE-221 00 Lund, Sweden}

\author{Richard R. Lane}
\affiliation{Instituto de Astrof\'isica, Pontificia Cat\'olica de Chile,
Av. Vicuna Mackenna 4860, 782-0436 Macul, Santiago, Chile}
\affiliation{Millennium Institute of Astrophysics, Av. Vicu\~na Mackenna 
4860, 782-0436 Macul, Santiago, Chile}

\author{Steven R. Majewski}
\affiliation{Department of Astronomy, University of Virginia, Charlottesville,
VA 22904, USA}

\author{Viktor Malanushenko}
\affiliation{Apache Point Observatory, P.O. Box 59, Sunspot, NM 88349, USA}

\author{Szabolcs~M{\'e}sz{\'a}ros}
\affiliation{ELTE E\"otv\"os Lor\'and University, Gothard
Astrophysical Observatory, Szombathely, Hungary}
\affiliation{Premium Postdoctoral Fellow of the Hungarian Academy of
Sciences}

\author{David L. Nidever}
\affiliation{Department of Physics, Montana State University, P.O. Box 173840, 
Bozeman, MT 59717-3840, USA}
\affiliation{National Optical Astronomy Observatory, 950 North Cherry Ave, 
Tucson, AZ 85719, USA}

\author{Christian Nitschelm}
\affiliation{Centro de Astronom\'ia (CITEVA), 
Universidad de Antofagasta, Avenida Angamos 601, Antofagasta 1270300, Chile}

\author{Kaike Pan}
\affiliation{Apache Point Observatory, P.O. Box 59, Sunspot, NM 88349, USA}

\author{Ricardo P. Schiavon}
\affiliation{Astrophysics Research Institute, Liverpool John Moores
University, Liverpool L3 5RF, UK}

\author{Donald P.\ Schneider}
\affiliation{Department of Astronomy and Astrophysics, The Pennsylvania 
State University, University Park, PA 16802}
\affiliation{Institute for Gravitation and the Cosmos, The Pennsylvania 
State University, University Park, PA 16802}

\author{John C. Wilson}
\affiliation{Department of Astronomy, University of Virginia, Charlottesville,
VA 22904, USA}

\author{Olga Zamora}
\affiliation{Instituto de Astrof\'isica de Canarias (IAC), E-38205 La Laguna, Tenerife, Spain}
\affiliation{Universidad de La Laguna, Dpto. Astrof\'isica, E-38206 La Laguna, Tenerife, Spain}

\begin{abstract} 
We map the trends of elemental abundance ratios across the Galactic disk,
spanning $R=3-15\kpc$ and midplane distance $|Z|=0-2\kpc$, for 15 elements
in a sample of 20,485 stars measured 
by the SDSS/APOGEE survey (O, Na, Mg, Al, Si, P, S, K, Ca, V, Cr, Mn, 
Fe, Co, Ni).  Adopting Mg rather than Fe as our reference element, and 
separating stars into two populations based on [Fe/Mg], we find that the median
trends of [X/Mg] vs.\ [Mg/H] in each population
are nearly independent of location in the Galaxy.
The full multi-element cartography can be summarized by combining these nearly
universal median sequences with our measured metallicity distribution functions
and the relative proportions of the low-[Fe/Mg] (high-$\alpha$) and 
high-[Fe/Mg] (low-$\alpha$) populations, which depend strongly on $R$ and $|Z|$.
We interpret the median 
sequences with a semi-empirical ``2-process'' model that describes both the
ratio of core collapse and Type Ia supernova contributions to each element and
the metallicity dependence of the supernova yields. These observationally
inferred trends can provide
strong tests of supernova nucleosynthesis calculations. Our results lead to
a relatively simple picture of abundance ratio variations in the Milky Way,
in which the trends at any location can be described as the sum of two
components with relative contributions that change systematically 
and smoothly across the Galaxy. Deviations from this picture and future 
extensions to other elements can provide further insights into the physics
of stellar nucleosynthesis and unusual events in the Galaxy's history.
\end{abstract}

\section{Introduction}

In the present-day Galaxy, iron-peak elements are produced in approximately
equal measure by core collapse supernovae (CCSN) and Type Ia supernovae
(SNIa).  Oxygen and magnesium production is dominated by CCSN, while heavier
$\alpha$-elements such as silicon, sulfur, and calcium are expected to have
significant contributions from SNIa (see, e.g.,
\citealt{Timmes1995,Kobayashi2006,Nomoto2013,Andrews2017,Rybizki2017}).
The distribution of stars
in the space of \afe-\feh\ has long been one of the crucial diagnostics of
chemical evolution, with \afe\ providing a rough chemical ``clock'' because of
the differing timescales of CCSN and SNIa enrichment 
\citep{McWilliam1997}.\footnote{We follow standard notation in which
  the bracketed abundance ratio 
  $[{\rm X}/{\rm Y}] = \log ({\rm X}/{\rm Y}) - \log({\rm X}/{\rm Y})_\odot$
  where $({\rm X}/{\rm Y})_\odot$ is the solar abundance ratio and
  $\log$ denotes a base-10 logarithm.
}
In the
solar vicinity this distribution is bimodal, with populations of
``high-$\alpha$'' and ``low-$\alpha$'' stars at sub-solar \feh.  The older,
high-$\alpha$ population is kinematically hotter and geometrically thicker,
allowing chemical separation of the the ``thick'' and ``thin'' disks
\citep{Fuhrmann1998,Prochaska2000,Bensby2003},
though this identification is blurred in the outer Galaxy by
disk flaring 
\citep{Minchev2015,Minchev2017,Bovy2016,Mackereth2017}.

The Apache Point Observatory Galactic Evolution Experiment (APOGEE;
\citealt{Majewski2017}) of the Sloan Digital Sky Survey (SDSS;
\citealt{York2000,Eisenstein2011,Blanton2017}) has extended the maps of the
\afe-\feh\ distribution over a large range of the Milky Way disk
(\citealt{Anders2014,Nidever2014}; \citealt{Hayden2015}, hereafter H15),
spanning $R=3-15\kpc$ and vertical height $|Z|=0-2\kpc$.  
With a large sample and precise, well characterized abundance measurements,
APOGEE data allow measurement of the intrinsic spread of \afe\ 
in the thin and thick disk populations at a given \feh\ \citep{Bertran2016}.
This paper extends ``chemical cartography'' of the Galactic disk to many of
the other elements measured by APOGEE: O, Na, Mg, Al, Si, P, S, K, Ca, V, Cr,
Mn, Co, and Ni.  These elements trace a variety of nucleosynthetic
pathways with different dependences on stellar mass and metallicity.
Their relative abundances can teach us about both the nucleosynthetic
processes themselves and the history of star formation and chemical
enrichment across the Galaxy.  

For practical reasons, previous studies of many elements for large
stellar samples have usually focused on the solar neighborhood
(e.g., \citealt{Timmes1995,Reddy2003,Reddy2006,Bensby2003,Bensby2005,
Bensby2014,Kobayashi2006,Adibekyan2012}).  These studies show
that the chemical dichotomy of the thick and thin disks can
be traced through many individual $\alpha$ elements, that trends
for iron-peak elements track those of iron as expected,
that neutron capture elements ($r$- and $s$-process) frequently
show distinctive behavior with metallicity and age, and that
multiple principal components are needed to represent the
distribution of local stars in the multi-dimensional abundance
space \citep{Ting2012,Andrews2017}.
Analogous to H15, this paper extends such analyses to span the
Galactic disk, taking advantage of APOGEE's much larger survey
volume and sample size.
We defer discussion of C and N to a separate paper 
(S. Hasselquist et al., submitted) because
their abundances in the evolved stars observed by APOGEE are strongly affected
by internal mixing on the red giant branch rather than reflecting birth
abundances.
One consequence of our choice is that the elements discussed in this paper are
ones whose production is likely dominated by CCSN and SNIa.
Future studies that combine APOGEE measurements with abundances from
optical surveys such as Gaia-ESO \citep{Gilmore2012}, 
LAMOST \citep{Luo2015}, and GALAH \citep{DeSilva2015}
can probe elements produced in intermediate mass stars or by exotic
mechanisms such as neutron star mergers.

H15's examination of metallicity distribution functions (MDFs) and
\afe\ ratios showed that the bimodality between 
high-$\alpha$ and low-$\alpha$ populations persists
across much of the disk, but that the relative fraction of stars in the two
populations depends strongly on Galactic location.  The high-$\alpha$ stars
trace a locus in \afe-\feh\ space that is nearly independent of location and
resembles the evolutionary track of simple chemical evolution models.  The
universality of this locus limits the variation of star formation efficiency
during the formation of the high-$\alpha$ population \citep{Nidever2014}.  
The distribution
of stars along the low-$\alpha$ locus shifts towards lower \feh\ at larger
radii, in accord with standard descriptions of the disk's stellar metallicity
gradient \citep{Cheng2012a}.  The skewness of the MDF changes
systematically with radius in the way expected if it has been shaped by radial
mixing of stars across the disk (\citealt{Schoenrich2009,Minchev2010}; H15).  
Most stars
in the inner Galaxy lie along the high-$\alpha$ sequence, but the distribution
of stars along this sequence shifts to higher \feh\ at lower $|Z|$ in a way
that suggests ``upside-down'' formation of the inner disk from a gas layer
that gets thinner over time \citep{Freudenburg2017}.
The fraction of high-$\alpha$ stars decreases drastically in the outer
Galaxy even at large $|Z|$, as suggested by earlier results
\citep{Bensby2011,Cheng2012b}.

For Mg and Fe abundance distributions, this paper confirms and further
quantifies the trends found in APOGEE data by 
\cite{Anders2014}, \cite{Nidever2014}, and H15 and in Gaia-ESO
data by \cite{Mikolaitis2014}.
For other elements we find that the ``cartography'' 
is relatively simple:
once we separate the high-$\alpha$ and low-$\alpha$ populations and
adopt Mg as our reference element, the median sequences of 
\xmg{X} vs.\ \mgh\ are nearly independent of Galactic location.
These sequences encode important constraints on supernova
nucleosynthesis, and they appear to do so in a way that is insensitive
to local variations in chemical enrichment history.
We interpret these sequences using a semi-empirical,
``2-process'' nucleosynthesis model that describes the element abundances in 
a given star as the sum of IMF-averaged CCSN and SNIa contributions
(IMF = initial mass function).
This model proves quite successful in describing our observed trends,
though this success and the position-independence of median sequences
may not extend to elements that have larger contributions from 
other enrichment channels.

After describing our data sample in \S\ref{sec:data}, we turn 
in \S\ref{sec:maps} to maps of abundance ratio trends in zones of
$R$ and $|Z|$.  Finding that these trends are nearly independent
of location, we examine them more closely 
using a high signal-to-noise ratio (SNR)
subset of the full disk sample in \S\ref{sec:trends}.
In \S\ref{sec:twoprocess} we define the 2-process model and
apply it to the interpretation of the observed median trends.
In \S\ref{sec:gradients} we present MDFs
and relative normalizations of the high- and low-$\alpha$ 
populations as a function of Galactic position, which in combination
with our abundance ratio trends provides an approximate description of
the full multi-element cartography for the elements examined here.
We summarize our
conclusions and discuss directions for future work in \S\ref{sec:conclusions}.

\section{Data}
\label{sec:data}

We use data from the fourteenth data release (DR14; \citealt{DR14})
of the SDSS/APOGEE survey \citep{Majewski2017}.
Targeting for the APOGEE survey is described by \cite{Zasowski2013},
\cite{Zasowski2017},
and the DR14 paper.  Roughly speaking, the APOGEE disk sample consists
of evolved stars with 2MASS \citep{Skrutskie2006} 
magnitudes $7 < H < 13.8$ sampled on a grid of sightlines 
accessible from the northern hemisphere at
Galactic latitudes $b=-8^\circ,-4^\circ,0^\circ,+4^\circ,+8^\circ$ 
over a wide range of Galactic longitudes.
$H$-band spectra are obtained with the 300-fiber APOGEE 
spectrograph (J. C. Wilson et al., submitted) 
on the Sloan Foundation 2.5m telescope
\citep{Gunn2006} at Apache Point Observatory.  \cite{Nidever2015}
describe the APOGEE data processing pipeline, which provides the input to
the APOGEE Stellar Parameters and Chemical Abundances Pipeline
(ASPCAP;\citealt{Garcia-Perez2016,Holtzman2015}), 
which fits effective temperatures,
surface gravities, and elemental abundances using a grid of synthetic
spectral models \citep{Meszaros2012,Zamora2015}
and a linelist described by \cite{Shetrone2015}.
Further details related to the DR14 data set, including the
empirical calibration of abundance scales, are described by
\cite{Holtzman2018}.
\cite{Jonsson2018} discuss comparisons between ASPCAP DR14 abundances
and measurements from optical spectra observed for the same stars
in other surveys.  We refer to results of these comparisons
at several points in the paper.

To minimize the
possibility of systematic issues with abundances with effective temperature
and/or surface gravity, we restrict the analysis to stars with 1$<$\logg$<$2,
roughly corresponding to $3700\K < T_{\rm eff} <4600\K$.
This surface gravity cut eliminates red clump (core helium burning) stars,
leaving only stars on the upper giant branch.
It also ensures that the stars in our sample can be observed by APOGEE 
over most of the distance range considered in this paper, minimizing
distance-dependent changes in the population being analyzed.
As data quality cuts, we require a signal-to-noise ratio of SNR$>80$ per 
pixel in the {\tt apStar} combined spectra ($\approx 0.22$\AA), 
and we require that no ``ASPCAP bad'' flags are set.
For the analyses in \S\ref{sec:trends} we apply a higher SNR threshold of 200.
Like H15, we focus on Galactic disk stars, with radial cuts
$3\kpc < R < 15\kpc$ and vertical cuts $|Z| < 2\kpc$.
We use spectrophotometric distance determinations similar 
to those used by H15; these compare well with other distance 
estimates, e.g., from \cite{Queiroz2018} and \cite{Wang2016}.
We use only stars targeted as part of the main APOGEE survey
(flag {\tt EXTRATARG=0}) to avoid any selection biases associated
with special target classes.
These cuts leave us with 20,485 stars, of which 13,350 have SNR$>200$.
 
\begin{figure} 
\includegraphics[width=0.45 \textwidth]{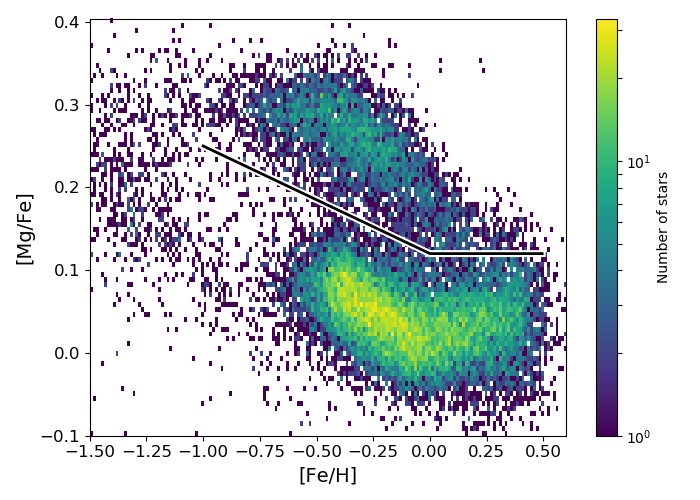}
\caption{Density in the $\mgfe$ vs.\ $\feh$ plane for
our full sample of 20,485 stars.
The black line marks our adopted division between
``high-$\alpha$'' and ``low-$\alpha$'' populations.  Recognizing
that the physical distinction between these populations is the amount
of iron enrichment from SNIa, we refer to them through most of the
paper as ``low-\femg'' and ``high-\femg,'' respectively.
}
\label{fig:MgFe}
\end{figure}

Figure~\ref{fig:MgFe} shows the distribution of our sample stars in the
familiar $\mgfe-\feh$ plane, displaying the usual bimodality of $\mgfe$
at sub-solar $\feh$.  As shown by H15 and below, the location of the
high-$\alpha$ and low-$\alpha$ sequences is nearly independent of
position within the Galactic disk, though the relative number of stars
on these sequences depends strongly on position.  The white line on
Figure~\ref{fig:MgFe} indicate the boundary we will use to separate
these two populations:
\begin{equation}
\begin{cases}
\mgfe > 0.12 - 0.13\feh, & \feh<0 \cr
\mgfe > 0.12,		     & \feh>0. \cr
\end{cases}
\label{eqn:boundary}
\end{equation}
While the two sequences converge at $\feh \ga 0$, making
the distinction of two populations ambiguous in this regime,
our boundary follows a shallow valley in the 2-d distribution,
and it is useful to separate high-metallicity stars with
different fractions of SNIa enrichment even if the distribution
is not clearly bimodal.

\section{Abundance ratio maps}
\label{sec:maps}

\subsection{Abundances relative to Mg}

Many studies of multi-element stellar abundances examine distributions of
\xfe{X}\ vs.  \feh.  Here we choose Mg as our reference element instead of Fe, 
so our basic diagrams are \xmg{X} vs.  \xh{Mg}.  
Because Mg is produced almost entirely
by CCSN, it is a simpler tracer of chemical enrichment than Fe, and it has
the same meaning (total enrichment from CCSN) on both the high-$\alpha$ and
low-$\alpha$ sequences.  
While oxygen would also be a suitable reference
element, we choose Mg because it is more robustly measured in APOGEE
(where O measurements come largely from OH and CO lines and 
are more sensitive to \teff) and because it is
more accessible to optical surveys.  
\cite{Jonsson2018} find that Mg is APOGEE's most accurately measured
$\alpha$ element relative to external measurements.
Other authors have used Mg
as a reference element in abundance ratio studies with similar
motivation (e.g., \citealt{Wheeler1989,Timmes1995,Fuhrmann1998,Cayrel2004}).  
In a similar spirit, we refer to
the high-$\alpha$ and low-$\alpha$ populations hereafter as
low-\femg\ and high-\femg, respectively, since the physical distinction
between them arises from the absence or prevalence of SNIa iron enrichment
rather than an enhancement of $\alpha$ elements relative to CCSN iron.
(One could more simply say low-iron and high-iron, but it is the
iron {\it relative} to $\alpha$-elements that matters for our
purposes, so we adopt the more specific terms to avoid confusion.)
We examine median abundance ratio trends separately for these two populations,
and we find that the combination of this practice and our choice of
reference element greatly simplifies the description of abundance
trends as a function of Galactic position.

\begin{figure*} 
\includegraphics[width=0.48 \textwidth]{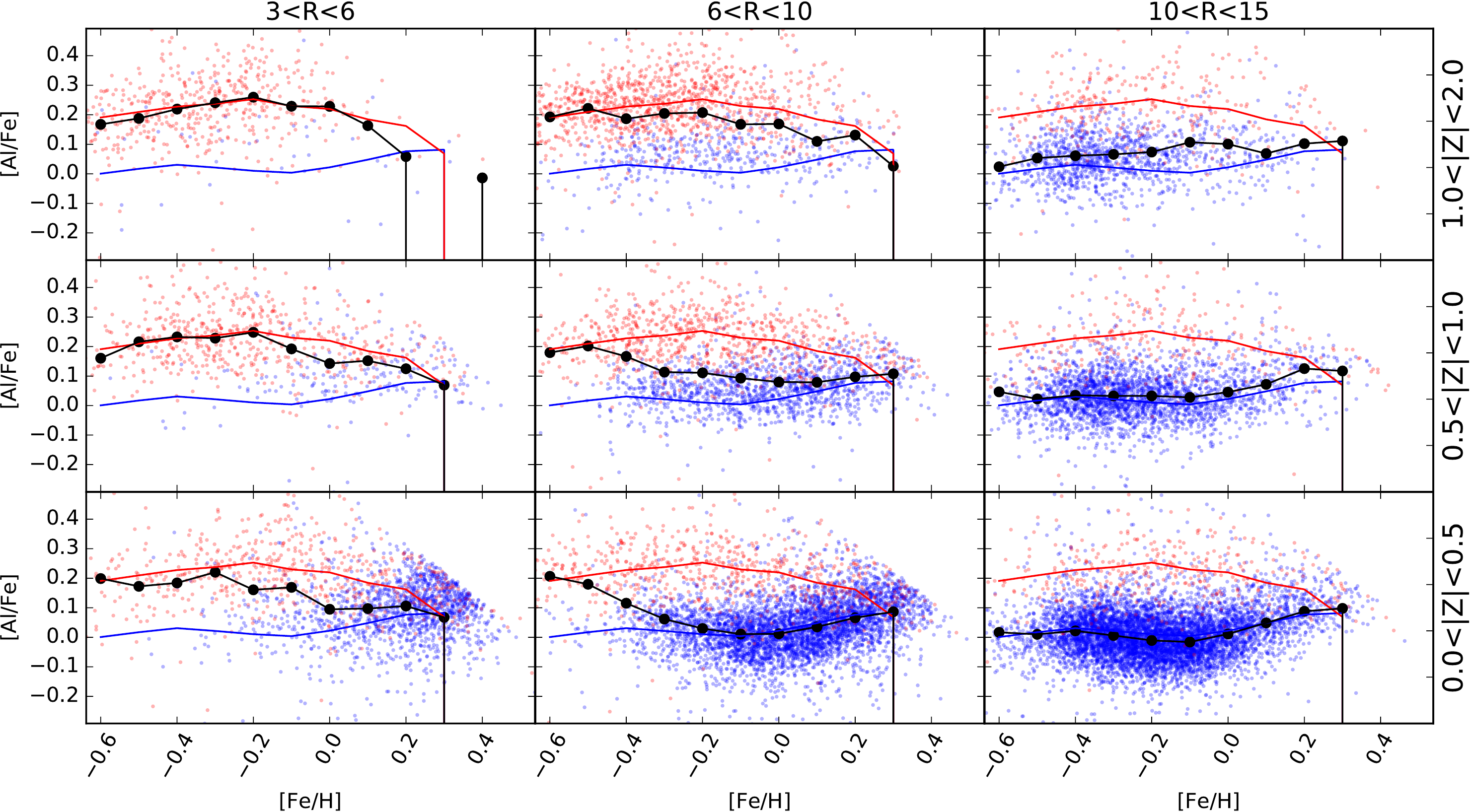}
\hskip 0.2truein
\includegraphics[width=0.48 \textwidth]{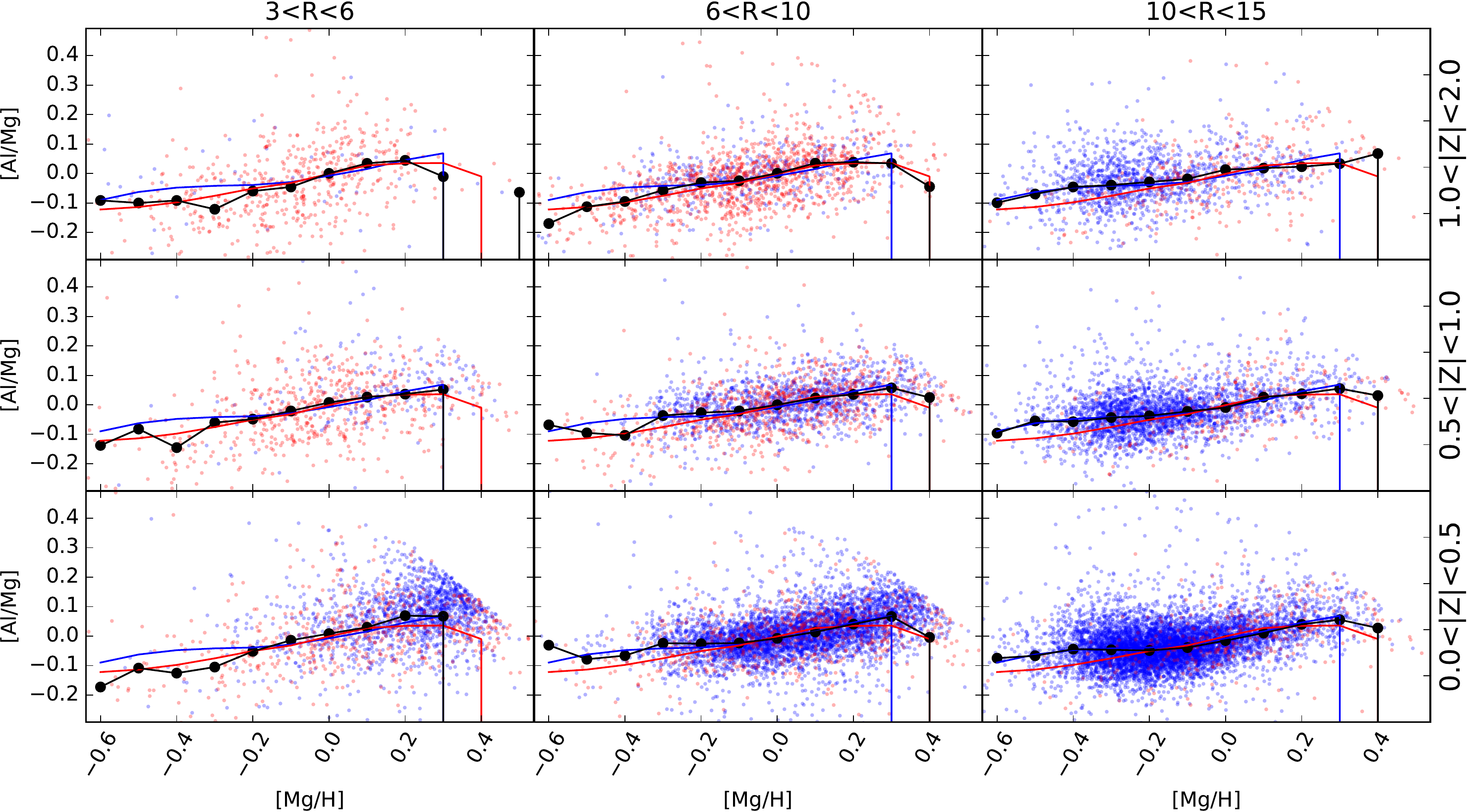}
\caption{
{\it (Left)} Distribution of sample stars in \xfe{Al}\ vs. \feh\ 
in nine Galactic zones, with radial ranges 
$R/\kpc = 3-6$ (left column), $6-10$ (middle column),
$10-15$ (right column) and vertical positions
$|Z|/\kpc = 0-0.5$ (bottom row), $0.5-1$ (middle row),
$1-2$ (top row).  
Red and blue points show stars in the low-\femg\ (high-$\alpha$)
and high-\femg (low-$\alpha$) populations, respectively, based on
the division marked in Figure~\ref{fig:MgFe}.  Large points connected by
the black solid line show the median \xfe{Al} in bins of \feh\ in each
Galactic zone.  Red and blue curves show the median trends for the
two populations in the full sample; these curves are the same in all panels.
{\it (Right)} Same as the left panels, but with Mg rather than Fe as
the reference element.  
}
\label{fig:AlFeMg}
\end{figure*}

Figure~\ref{fig:AlFeMg} illustrates this simplification for the case
of Al. The left half of the figure shows \xfe{Al}\ vs.\ \feh\ for all
sample stars in nine zones of $R,|Z|$, with red and blue points 
marking stars in the low-\femg\ (high-$\alpha$) and 
high-\femg\ (low-$\alpha$) populations, respectively.
Large points connected by the black line show the median abundance
ratio of all stars in bins of \feh.  
This median trend changes shape with Galactic position, and its shape
has no obvious physical interpretation.
The scatter about the median trend is large at all locations.
However, the median trends for the low-\femg\ and high-\femg\
populations individually (red and blue curves) are nearly
independent of location, and the scatter about the median
trend within each population is fairly small.  The changing
shape of full sample median reflects the changing ratio of
low-\femg\ to high-\femg\ stars as a function of Galactic
position and metallicity.
Put simply, thin-disk stars have low
\xfe{Al} at a given \feh\ because they have extra Fe from SNIa
enrichment, and the median \xfe{Al} vs.\ \feh\ is therefore
shaped by the ratio of thin-disk to thick-disk stars.

The right half of the figure plots \xmg{Al}\ vs. \mgh\ for the
same stars in the same zones.
With Mg rather than Fe as reference element, the median trends
are nearly identical for
low-\femg\ and high-\femg\ stars, which is expected because Al 
production is dominated by CCSN and thus unaffected by the
presence of SNIa iron.  
The full sample median, shown by the large points and black line,
is nearly independent of position.
The trend is physically sensible for an odd-$Z$ element such as Al:
the \xmg{Al} ratio rises with increasing \mgh\ because the yield
of odd-$Z$ elements increases with metallicity.
Some other elements exhibit
significantly different trends in the two populations because 
of differing SNIa contributions, but the near independence
of Galactic position continues to hold.

\subsection{Dual chemical sequences in [Fe/Mg]}
\label{sect:dual}

\begin{figure*} 
\begin{center}
   \includegraphics[width=0.9 \textwidth]{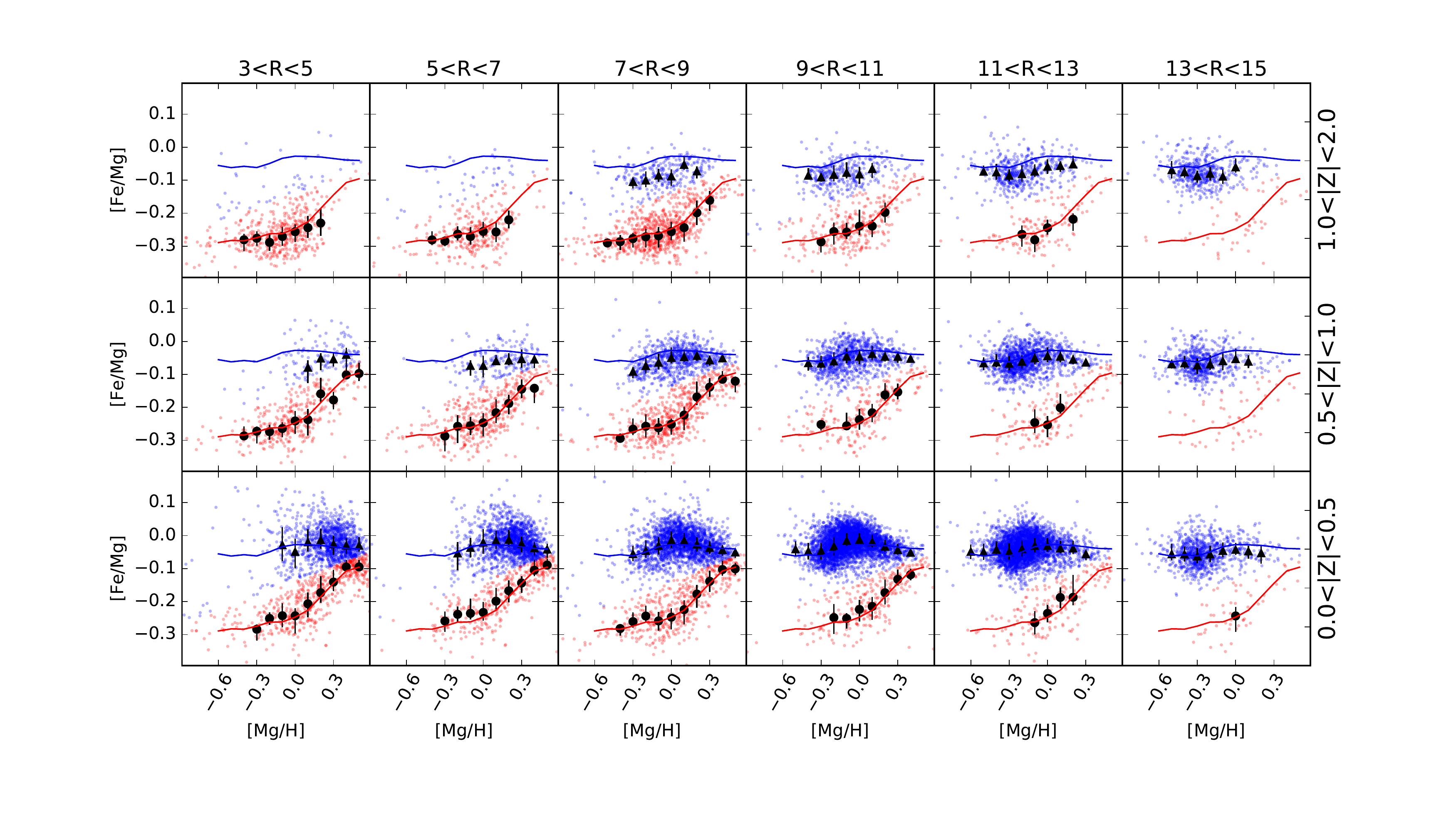}
\end{center}
\caption{\xmg{Fe} vs \xh{Mg} for 18 Galactic zones:
from left to right, $R/\kpc = $3-5, 5-7, 7-9, 9-11, 11-13, 13-15;
from top to bottom, $|Z|/\kpc = $1-2, 0.5-1, 0-0.5.
Red and blue points show stars in the
low-\femg\ and high-\femg\ populations, respectively.
Large black points show the median \femg\ in 0.1-dex bins of
\mgh\ for each population in each zone, with triangles showing the high-\femg\ population and circles the low-\femg\ population;
red and blue curves show corresponding median
relations for the full sample.  
For each zone, points are plotted only if an \mgh\ bin has at least 15
stars.  Error bars denote the interquartile range of \femg\ for the
stars in each bin.  The statistical errors in the median are much
smaller than this range (by $\sim \sqrt{N_{\rm stars}}$).
}
\label{fig:FeMg}
\end{figure*}

Figure~\ref{fig:FeMg} plots the distribution of stars in \xmg{Fe} and \xh{Mg},
now in six radial zones and the same three vertical zones.  
Small points show stars in the low-\femg\ population
(red) and high-\femg\ population (blue).
In each zone, large points show
the median \femg\ in bins of \mgh\ for the two populations.  Points
are plotted only if a bin contains at least 15 stars in
a given population.  Red and blue curves (the same in each panel)
show median relations for the entire sample.

This figure is similar to figure 4 of H15, but our choice
of Mg as reference element inverts the $y$-axis and rescales the $x$-axis.
Numerous studies show that the low-\femg\ population is more prominent
at small $R$ and large $|Z|$ --- i.e., the chemically defined thick disk has
a smaller scale length and larger scale height 
(\citealt{Bensby2011,Bovy2012,Cheng2012b,Mikolaitis2014,Nidever2014}; H15).
In the high-\femg\ (thin-disk) population, the mode of \mgh\ 
shifts from roughly $+0.3$ in the inner Galaxy to roughly $-0.3$ in
the outer Galaxy (see Fig.~\ref{fig:mdfstats} below), 
reflecting the stellar metallicity gradient.
Radial trends in the low-\femg\ population are weaker, though 
in the inner disk the $\mgh$ distribution shifts to higher
metallicity with decreasing $|Z|$.  Most importantly from the
point of view this paper, the median trends of $\femg$ with $\mgh$
for the two populations
are nearly constant throughout the disk, as shown by the close
agreement between the points in each zone and the corresponding
curves.  This constancy of the low-\femg\ and high-\femg\ locus
confirms similar findings by \cite{Nidever2014} and H15 with
the larger sample and improved measurements of the DR14 data set.
There are small shifts in the median locus of high-\femg\ stars
at $|Z|=1-2\kpc$, a point we discuss further in \S\ref{sec:offsets}
below.

\begin{figure*} 
\begin{center}
   \includegraphics[width=0.9 \textwidth]{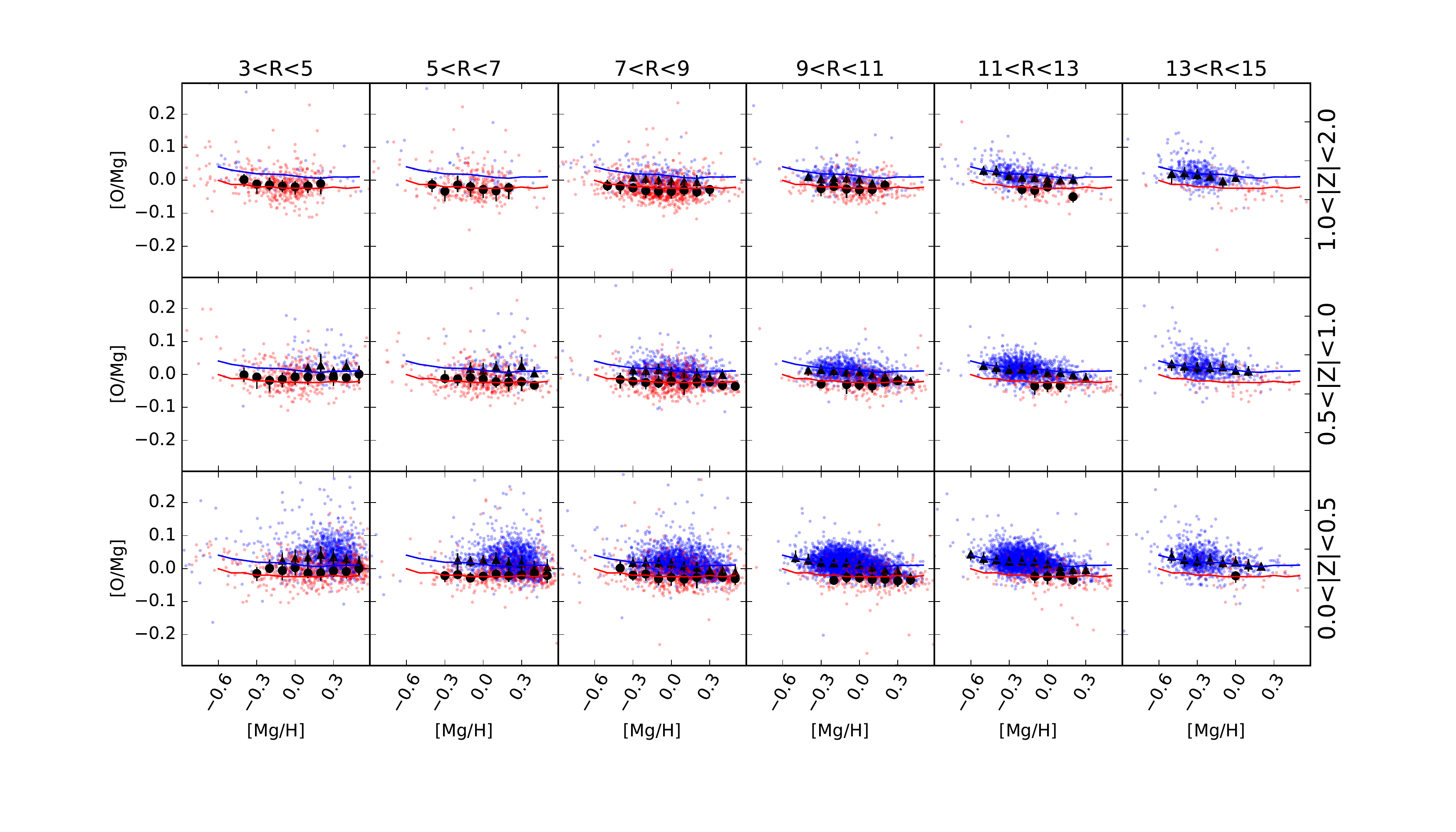}
\end{center}
\caption{Same as Figure~\ref{fig:FeMg}, but for \xmg{O}.
}
\label{fig:OMg}
\end{figure*}

\begin{figure*} 
\begin{center}
   \includegraphics[width=0.9 \textwidth]{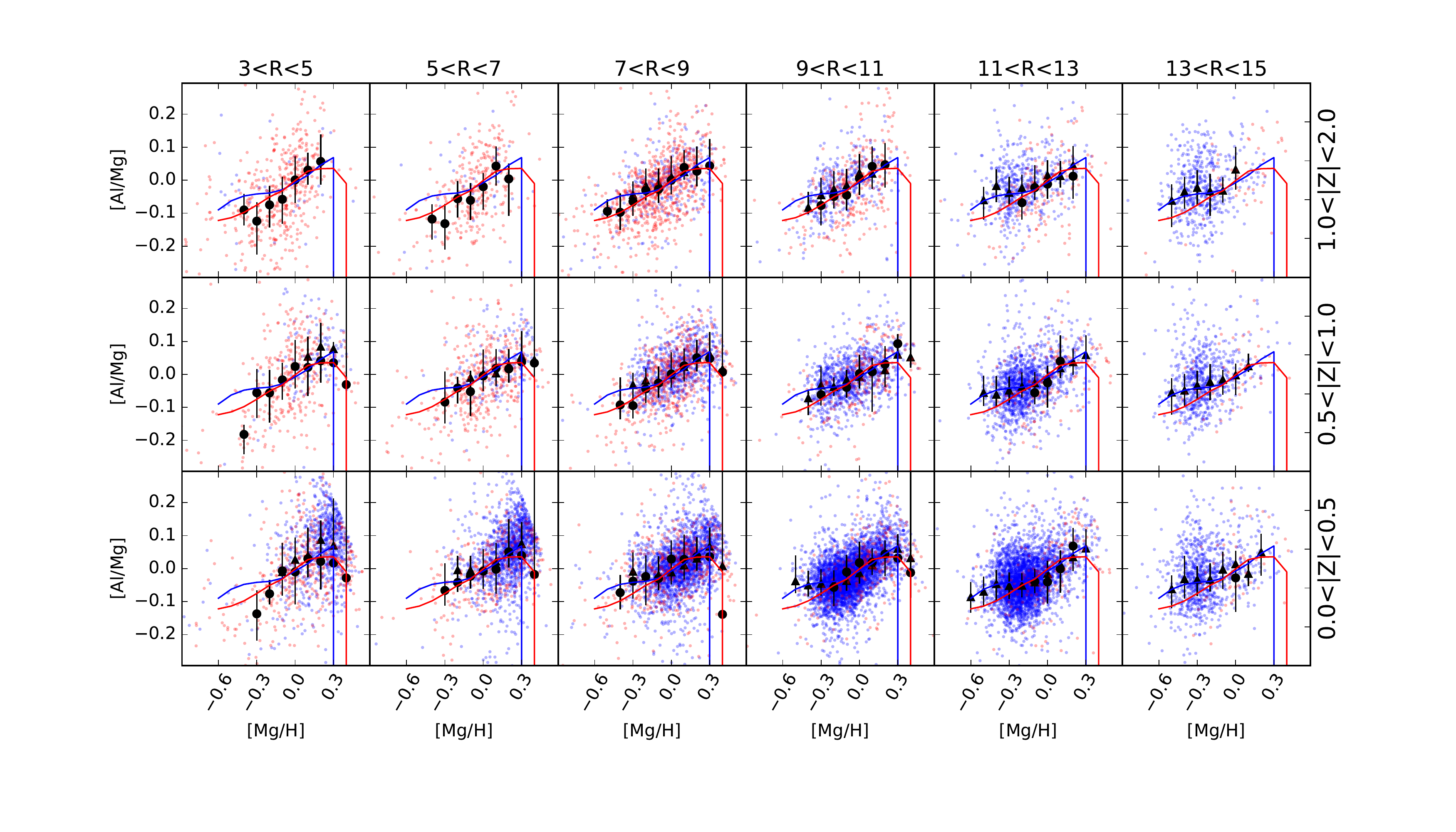}
\end{center}
\caption{Same as Figure~\ref{fig:FeMg}, but for \xmg{Al}.
}
\label{fig:AlMg}
\end{figure*}

\begin{figure*} 
\begin{center}
    \includegraphics[width=0.9 \textwidth]{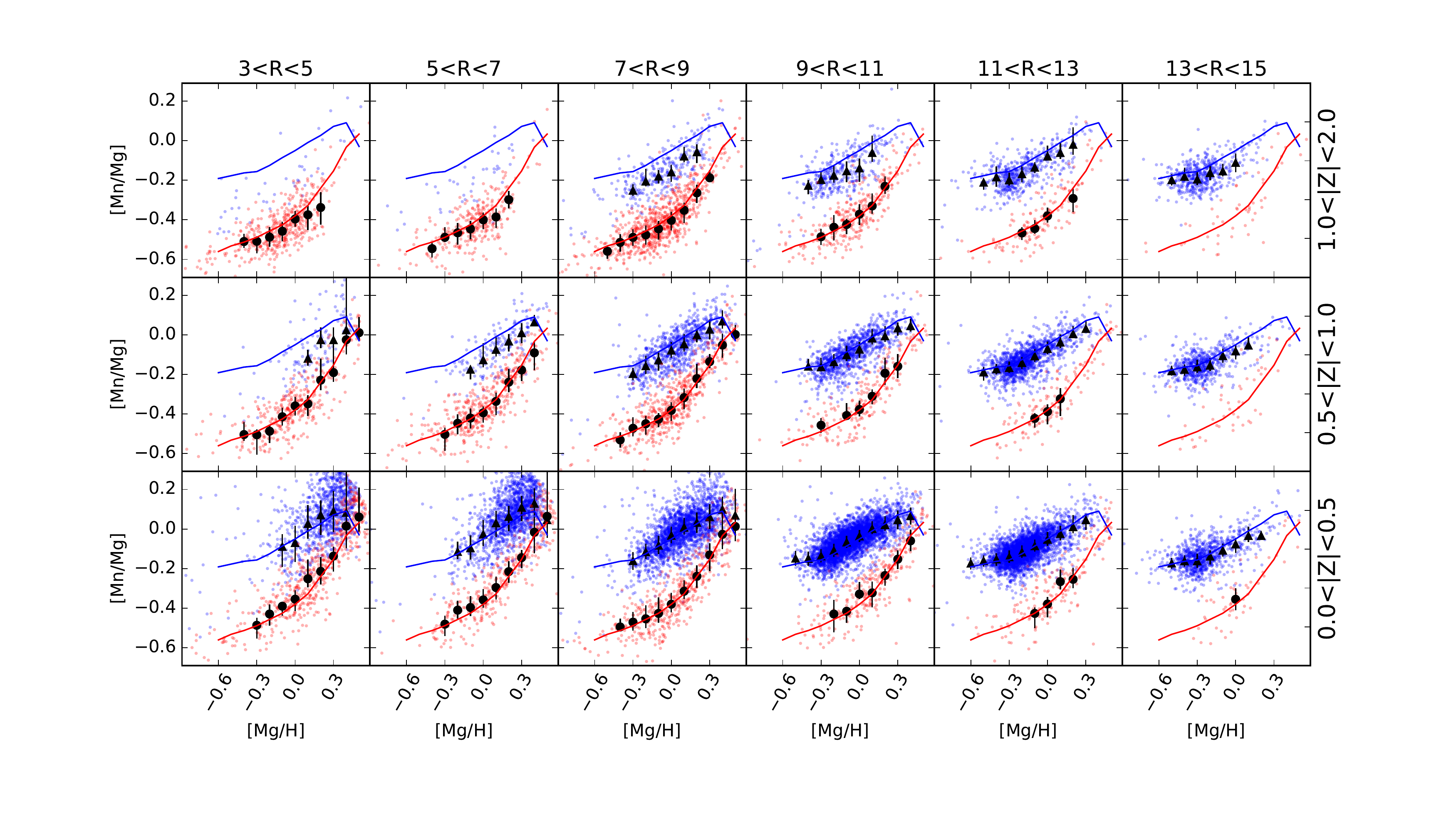}
\end{center}
\caption{Same as Figure~\ref{fig:FeMg}, but for \xmg{Mn}.
}
\label{fig:MnMg}
\end{figure*}

Figure~\ref{fig:OMg} shows a similar map of abundance ratio trends
for \xmg{O} vs.\ \mgh.  Like Mg, O is an $\alpha$-element whose
production should be dominated by CCSN.  The behavior is remarkably
simple: for both low-\femg\ and high-\femg\ populations, median 
trends are nearly flat at $\xmg{O}=0$ at all Galactic positions.
This flatness indicates that IMF-averaged yields of 
Mg and O are independent of metallicity over the range investigated here,
or at least that any metallicity dependence is the same for the two 
elements.
In the full-sample median, the trend for low-\femg\ stars is
offset from that of high-\femg\ stars by about $-0.02\,$dex in
\xmg{O}.  We cannot be certain that this small offset is not an artifact
of abundance measurement systematics that differ in the two populations. 
The abundance of oxygen is determined largely from OH lines and its value 
is sensitive to \teff; \cite{Jonsson2018} discuss possible
metallicity dependent systematics in \teff.
However, we have checked that the distributions of $T_{\rm eff}$ and
$\log g$ are nearly identical for the low-\femg\ and high-\femg\
subsets of our sample, so there is no obvious reason for
\teff-dependent systematics to produce an offset in \xmg{O}
between the populations.

Figure~\ref{fig:AlMg} shows results for \xmg{Al} in the same format,
while Figure~\ref{fig:MnMg} shows results for Mn, an odd-$Z$ 
element near the iron peak.  As seen previously in Figure~\ref{fig:AlFeMg},
the median trends for Al show an increasing \xmg{Al} towards higher
\mgh\ as expected for an element whose yield increases with metallicity.
The trends for low-\femg\ and high-\femg\ stars are nearly identical,
consistent with Al having no significant SNIa contribution, and
they are independent of Galactic position within the statistical
noise of our sample.  For Mn, on the other hand, the metallicity
dependence is stronger and the trends for low-\femg\ 
and high-\femg\ stars are offset and have different slopes.
The higher \xmg{Mn} in high-\femg\ stars is a sign that 
a substantial fraction of Mn is produced in SNIa, again consistent
with theoretical expectations (see further discussion below).
The trends within each population are again nearly independent
of Galactic position.  At high $|Z|$ the \xmg{Mn} trend of the high-\femg\ 
population lies below the global median trend by 0.05-0.1 dex,
a larger difference than seen for Fe, O, or Al, but small compared
to the separation from the low-\femg\ trend.

For the other elements considered, median trends are also
nearly independent of Galactic position.  
In \S\ref{sec:offsets} we will
discuss the small deviations from this universal behavior,
but we turn first to a discussion and characterization of the global 
median trends and their interpretation.

\section{Global trends and their implications}
\label{sec:trends}

Given the approximate constancy of the median sequences, we investigate the
nucleosynthetic trends by combining all of the data from different zones into
single diagrams.  
To maximize the accuracy, we restrict this combined sample to
stars with ${\rm SNR} > 200$ per pixel; 
an advantage of large samples is the ability
to choose stars with the highest quality data.
This high-SNR subsample still spans the full range of $R$ and $|Z|$,
though relative to the full sample it has a smaller fraction of
stars at small and large $R$.
As an additional check for systematic effects in the abundance
measurements, we have examined
the trends for the coolest stars in our high-SNR subsample
($T_{\rm eff} < 4000\K$, approximately 13\% of the total)
and for the hottest stars
($T_{\rm eff} > 4400\K$, approximately 9\% of the total).
In general we find no significant difference in global trends
for these coolest and hottest subsets, with the exception of
Al and V, discussed in \S\ref{sec:trends_oddz} and \S\ref{sec:trends_peak},
respectively.

\begin{figure*} 
\begin{center}
\includegraphics[width=0.9\textwidth]{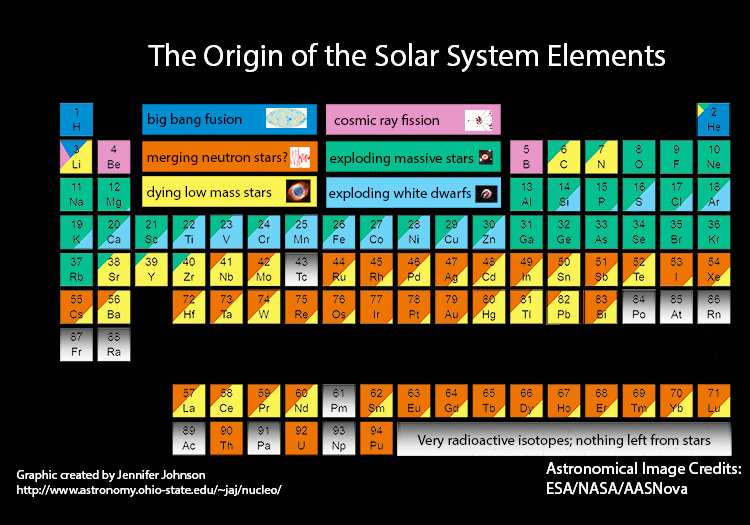}
\end{center}
\caption{Periodic table of elements color-coded by nucleosynthetic origin,
for solar system abundance ratios.
Green and light blue represent contributions from CCSN and SNIa, respectively,
which dominate the production of elements considered in this paper.
Yellow represents contribution from AGB stars.  Heavy $r$-process elements
are here attributed to merging neutron stars, though the extent to
which these dominate $r$-process production remains uncertain.
Elements in gray have short radioactive decay times, so terrestrial
incidence of these elements does not originate in stellar sources.
B, Be, and some Li are produced by cosmic ray spallation.
Essentially all H, most He, and some Li originate in the hot early 
universe.
}
\label{fig:periodic}
\end{figure*}

Our discussion of nucleosynthetic sources for different elements is
based on that of \citeauthor{Andrews2017} (\citeyear{Andrews2017},
see their \S 4.2 and Appendix B), 
which is in turn derived from CCSN yields of
\cite{Chieffi2004} and \cite{Limongi2006}, SNIa yields from the
W70 model of \cite{Iwamoto1999}, and AGB yields from \cite{Karakas2010}.
As a useful intuitive reference, we show in 
Figure~\ref{fig:periodic} a version of the periodic table
constructed by J.\ Johnson and I.\ Ivans, in which individual
elements are color-coded according to their production 
mechanisms.\footnote{For alternative forms of this graphic and a link
  to its Creative Commons license, see
{\tt http://www.astronomy.ohio-state.edu/$\sim$jaj/nucleo}.}
This coding is informed by both theoretical predictions and
empirical data, but it is necessarily uncertain; for example,
the attribution of heavy $r$-process elements to neutron star 
mergers remains a conjecture, strengthened by the spectroscopic
observations of the neutron star merger event GW170817 \citep{Pian2017}.
For the elements examined in this paper, one can see from the
figure that O, Na, Mg, and Al are thought to originate almost
entirely from CCSN, that Si, P, S, K, and Ca are thought to
come predominantly from CCSN but with significant SNIa contributions,
and that V, Cr, Mn, Fe, Co, and Ni are thought to come at least 50\%
from SNIa, though with significant CCSN contributions in each case.
These characterizations apply to solar system abundances;
at low metallicity, the relative contributions of SNIa would be lower,
and contributions could be different in a galaxy
or stellar population with a different IMF or a radically different
star formation history.
The more detailed calculations presented by \cite{Andrews2017}
present a similar picture and show predictions as a function of \feh.
Figure~13 of \cite{Rybizki2017} is another useful reference for
relative CCSN, SNIa, and AGB contributions to different elements,
with two different yield sets.

\subsection{$\alpha$-elements}
\label{sec:trends_alpha}

\begin{figure} 
\includegraphics[width=0.45 \textwidth]{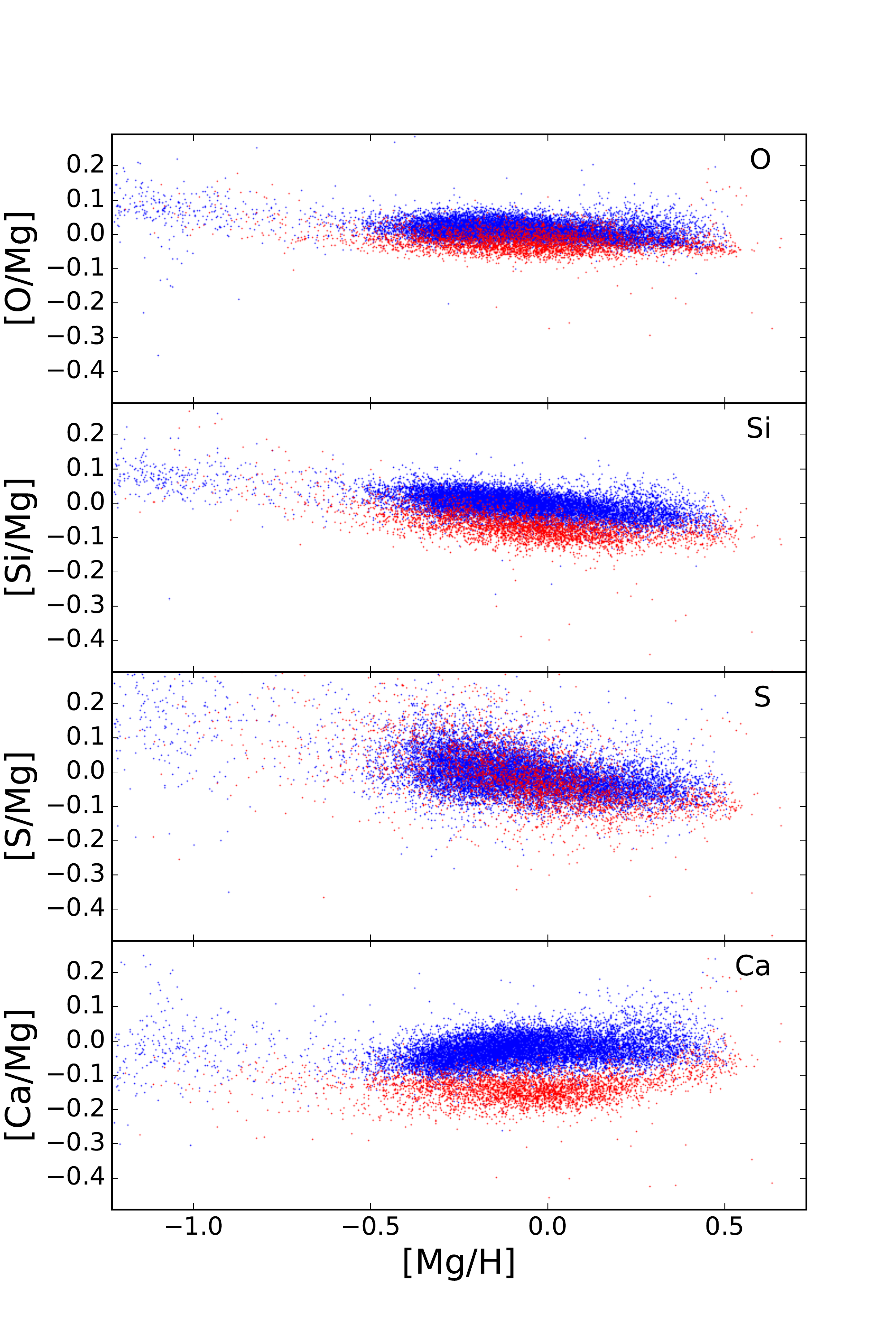} 
\caption{\xmg{X} vs. \mgh\ for $\alpha$-elements, for all sample stars with
${\rm SNR}>200$.  Red and blue points show the low-\femg\ and high-\femg\
populations, containing 2038 and 11,311 stars, respectively.
}
\label{fig:alpha_2alpha} 
\end{figure}

Figure~\ref{fig:alpha_2alpha} plots \xmg{X} vs. \mgh\ for the
$\alpha$-elements O, Si, S, and Ca.  For an element produced
purely by CCSN with metallicity-independent IMF-averaged yield,
the \xmg{X} ratio is expected to be independent of \mgh, and it should be
the same for low-\femg\ and high-\femg\ stars
(see further discussion at the end of \S\ref{sec:twoprocess_metdepyield}).
This is nearly the case for O, though there is a slight
offset between the two sequences and a slightly elevated
\xmg{O} at low metallicities.  For Si the metallicity trend
is stronger and more continuous, and the separation of the
two sequences is larger, consistent with the moderate 
SNIa contribution expected for Si.
However, while the expected SNIa contribution 
(relative to CCSN) is larger for S than for Si,
we find no separation between the two \xmg{S} sequences.
Ca is expected to have the largest SNIa contribution among
these elements, and it shows the largest separation between the low-\femg\
and high-\femg\ populations.  We quantify the relative CCSN vs.\ SNIa
contributions and the metallicity dependence of these contributions
in \S\ref{sec:twoprocess} below.

The possibility of systematics in APOGEE oxygen 
measurements was mentioned in \S \ref{sect:dual}. 
For Si, S, and Ca, \citet{Jonsson2018} show that the
APOGEE measurements are in fairly good agreement with independent 
estimates from optical spectra of the same stars,
and that the global loci of \xfe{X} vs \feh\  for these elements 
are in generally good agreement with loci determined independently 
in local samples.  

\begin{figure} 
\includegraphics[width=0.45 \textwidth]{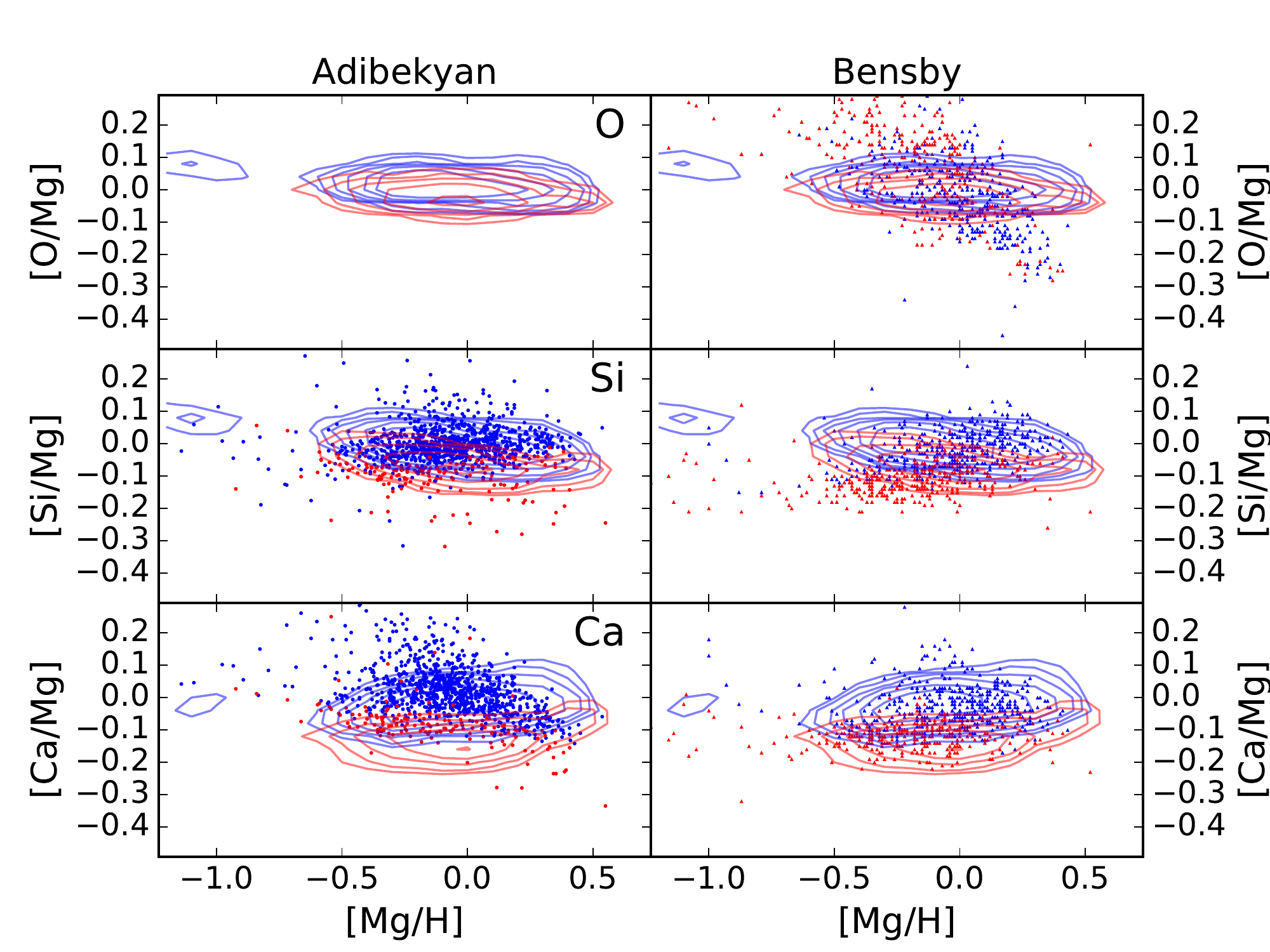} 
\caption{Comparison of APOGEE loci in \xmg{X} vs.\ \mgh\ to 
optical abundance measurements of solar neighborhood stars from
\cite{Adibekyan2012} and \cite{Bensby2014}, for O, Si, and Ca.
Logarithmically spaced
red and blue contours show the density of low-\femg\ and high-\femg\
stars, respectively, from the APOGEE disk sample.
Points show individual stars from the optical samples, with
the same color coding.  
The \cite{Adibekyan2012} data set does not have O abundances, and neither
data set has S abundances.
}
\label{fig:alpha_comparison} 
\end{figure}

For a more direct comparison to our \xmg{X}-\mgh\ distributions,
we have used the samples of \cite{Adibekyan2012} and \cite{Bensby2014}
and applied the \mgfe\ criterion
of Equation~(\ref{eqn:boundary}) to define two populations.
Figure~\ref{fig:alpha_comparison} plots individual stars
from the two optical samples over contours that
show the density of stars in our full APOGEE DR14 disk sample.
In contrast to the \cite{Jonsson2018} star-by-star
comparisons, the stars in the optical samples are different
from those in the APOGEE data.  The \cite{Adibekyan2012} data
do not include oxygen abundances, and neither data set includes
sulfur abundances.

The \cite{Bensby2014} data show a sloped trend of \xmg{O} vs.\ \mgh,
in contrast to the flat trends found in the APOGEE data set.
Given the challenges of inferring oxygen abundances from 
either optical spectra (principally non-LTE corrections) or
near-IR spectra (principally the impact of \teff\ on OH 
molecular abundances), we cannot be sure which of these
trends is more accurate.\footnote{``Non-LTE corrections'' refer
  to corrections to inferred abundances that account for departures
from local thermodynamic equilibrium (LTE) in stellar atmospheres.}
The APOGEE trend is much easier to 
understand physically, since Mg and O are both produced mainly
during hydrostatic evolution in high mass progenitors of CCSN,
and it is not clear what mechanism could create a
metallicity-dependent \xmg{O} ratio.

For \xmg{Si} the APOGEE data are in good agreement with
the \cite{Adibekyan2012} sample and fair agreement with the
\cite{Bensby2014} sample.  Without separating the two populations,
either of the optical samples might suggest a rising trend of
\xmg{Si} with \mgh, but the APOGEE data show that the trends
are slightly falling within each population.
In the solar neighborhood optical samples, the low-\femg\ 
stars are preferentially lower \mgh, leading to the net
rising trend. 
A trend of \xfe{Si} vs.\ \feh\ would be further complicated
by the differing \xmg{Fe} ratios of the two populations.

For \xmg{Ca} the APOGEE data are in good agreement with 
the \cite{Bensby2014} data.  The \cite{Adibekyan2012} data show slightly
higher \xmg{Ca} for the low-\femg\ population and an
extension to high \xmg{Ca} for lower metallicity stars
in the high-\femg\ population.

\subsection{Light odd-$Z$ elements}
\label{sec:trends_oddz}

\begin{figure} 
\includegraphics[width=0.45 \textwidth]{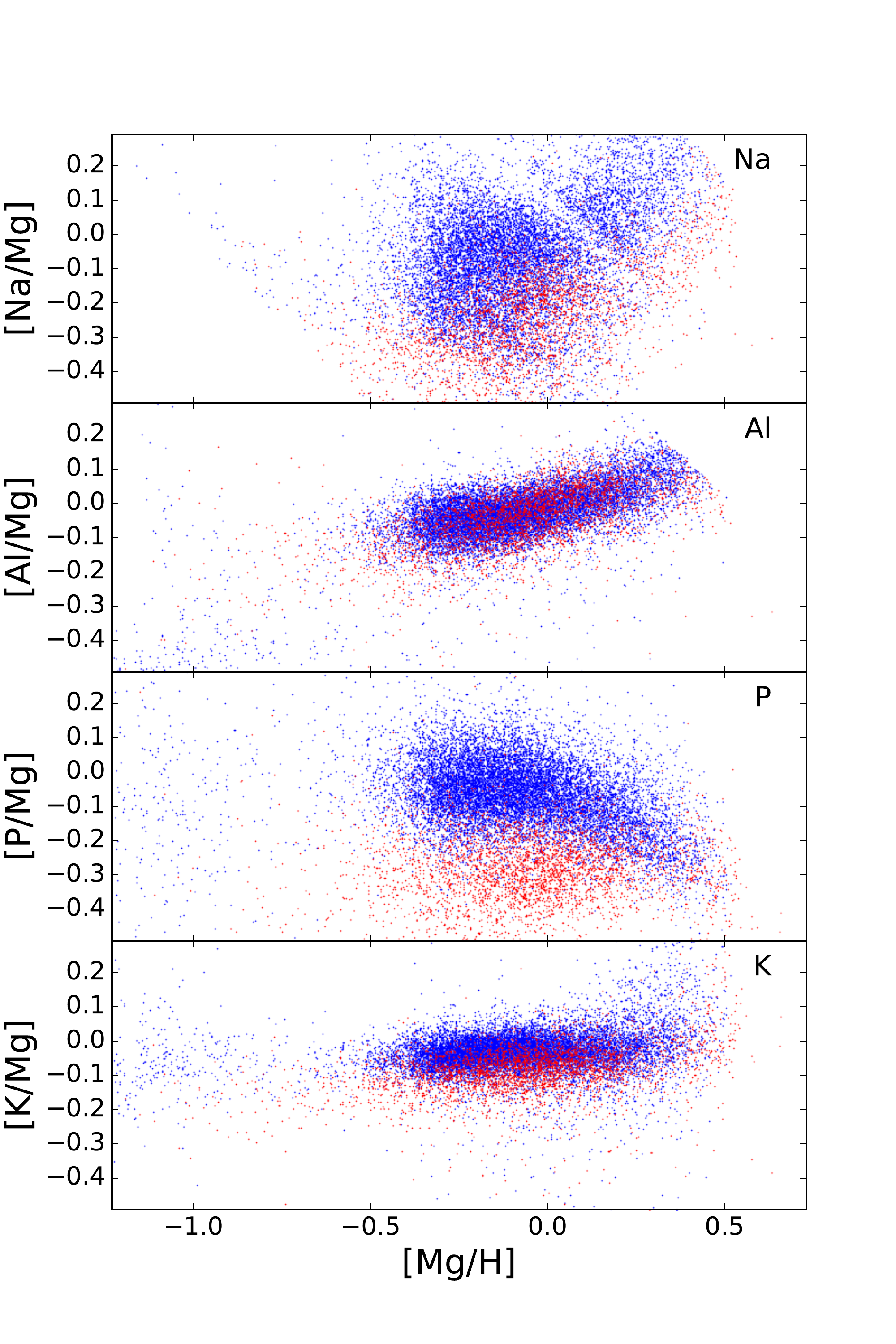} 
\caption{Same as Figure~\ref{fig:alpha_2alpha} for light odd-$Z$ elements.
}
\label{fig:oddz_2alpha} 
\end{figure}

Figure~\ref{fig:oddz_2alpha} plots \xmg{X} vs. \mgh\ for 
Na, Al, P, and K, elements with odd atomic number whose
production is dominated by CCSN.  Predicted yields for
these light odd-$Z$ elements increase with stellar metallicity
because their production requires a neutron excess, and
this excess requires elements heavier than H and He
\citep{Truran1971,Timmes1995,Woosley2002}.
Specifically, higher initial CNO abundances in CCSN progenitors
lead to more $^{14}{\rm N}$ production during hydrogen burning, the 
$^{14}{\rm N}$ is converted to $^{22}{\rm Ne}$ during helium burning,
and destruction of $^{22}{\rm Ne}$ by $\alpha$-captures becomes
a source of free neutrons.

Our results for Al are in qualitative agreement with expectations,
with \xmg{Al} increasing with \mgh\ and similar trends for
low-\femg\ and high-\femg\ populations.  
Trends for \xmg{K} are nearly flat, though slightly increasing.
For both Na and P we find large differences in \xmg{X} for
the low-\femg\ and high-\femg\ populations, counter to expectations
if production of these elements is dominated by CCSN.
We caution that our typical abundance errors for these 
elements are relatively large even for these ${\rm SNR} > 200$
spectra, ranging from 0.05 to 0.2 dex for Na 
and 0.05 to 0.1 dex for P 
depending on \teff\ and \mh.

There is an obvious artifact in the $\xmg{Na}-\mgh$ diagram:
a diagonal gap that corresponds to a narrow but sharp minimum
in the \xh{Na} distribution at $\xh{Na} \approx 0-0.1$, which
is most pronounced for stars with $\log g<1.5$.
This feature is not present in APOGEE DR12 or DR13 \xh{Na} abundances;
stars with $\xh{Na}$ in this range in DR12 or DR13 are assigned
systematically higher $\xh{Na}$ in DR14.
Despite an extensive investigation, we do not fully understand the origin
of this artifact, but it highlights the fact that APOGEE sodium
abundance measurements rely on two fairly weak lines, one of
which is significantly blended with a CO feature.

\begin{figure} 
\includegraphics[width=0.45 \textwidth]{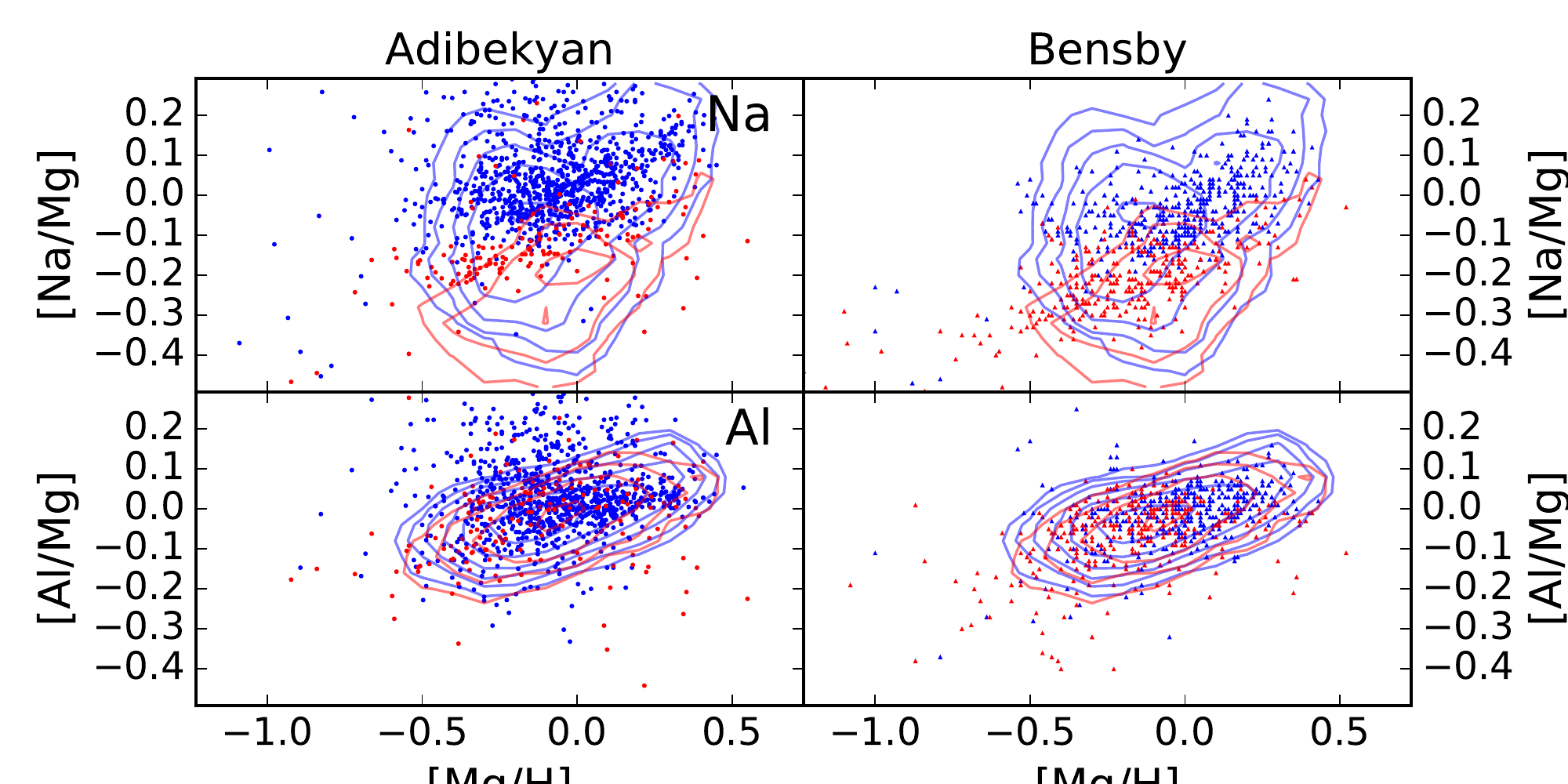} 
\caption{Same as Figure~\ref{fig:alpha_comparison} but for Na and Al.
  The \cite{Adibekyan2012} and \cite{Bensby2014} data sets do not include
P or K measurements.
}
\label{fig:oddz_comparison} 
\end{figure}

As discussed by \citet{Jonsson2018}, the APOGEE measurements for Na and Al 
are generally in good agreement with independent optical measurements
for the same stars, and the loci in \xfe{Na}-\feh\ and \xfe{Al}-\feh\
are similar to those found in local studies. 
Figure~\ref{fig:oddz_comparison} shows \xmg{Na} and \xmg{Al}
comparisons to the \cite{Adibekyan2012} and \cite{Bensby2014} samples, 
indicating reasonable agreement, somewhat better for
\cite{Bensby2014} than for \cite{Adibekyan2012}.
P is relatively unstudied, so there is little in the way of 
external validation. For K, \cite{Jonsson2018} find
substantial systematic differences with measurements made from optical 
spectra, so our K results should be interpreted with caution.

For Al specifically, we find that
the coolest stars in our sample ($T_{\rm eff}<4000\K$)
imply a steeper trend of \xmg{Al} vs.\ \mgh; the median \xmg{Al} of
these stars matches that of the full sample near $\mgh=0$, but it 
is slightly higher at high-\mgh\ and lower by about 0.05 dex
at $\mgh \approx -0.3$.  The visually estimated 
trend slope through the main locus
of points is about 0.20 for the full sample and about 0.35 for
the coolest stars (precise numbers depend on binning and trimming
of outliers, especially for the smaller cool, subset).
The hottest stars ($T_{\rm eff}>4400\K$) have slightly higher
median \xmg{Al} (by about 0.05 dex) at $\mgh \approx -0.3$; it is
difficult to translate this difference into a trend slope because
there are few hot stars in our sample with $\mgh > 0$.
Neither the coolest nor hottest subsamples show an offset 
of \xmg{Al} between the low-\femg\ and high-\femg\ populations,
so the inference that Al production is dominated by CCSN
remains robust.

For the other $\alpha$-elements and
light odd-$Z$ elements, we find no clear differences in median
trends for the coolest or hottest stars in the sample, within
the limits set by sample size and scatter in the abundance ratios.

\subsection{Iron-peak elements}
\label{sec:trends_peak}

\begin{figure}
\includegraphics[width=0.45 \textwidth]{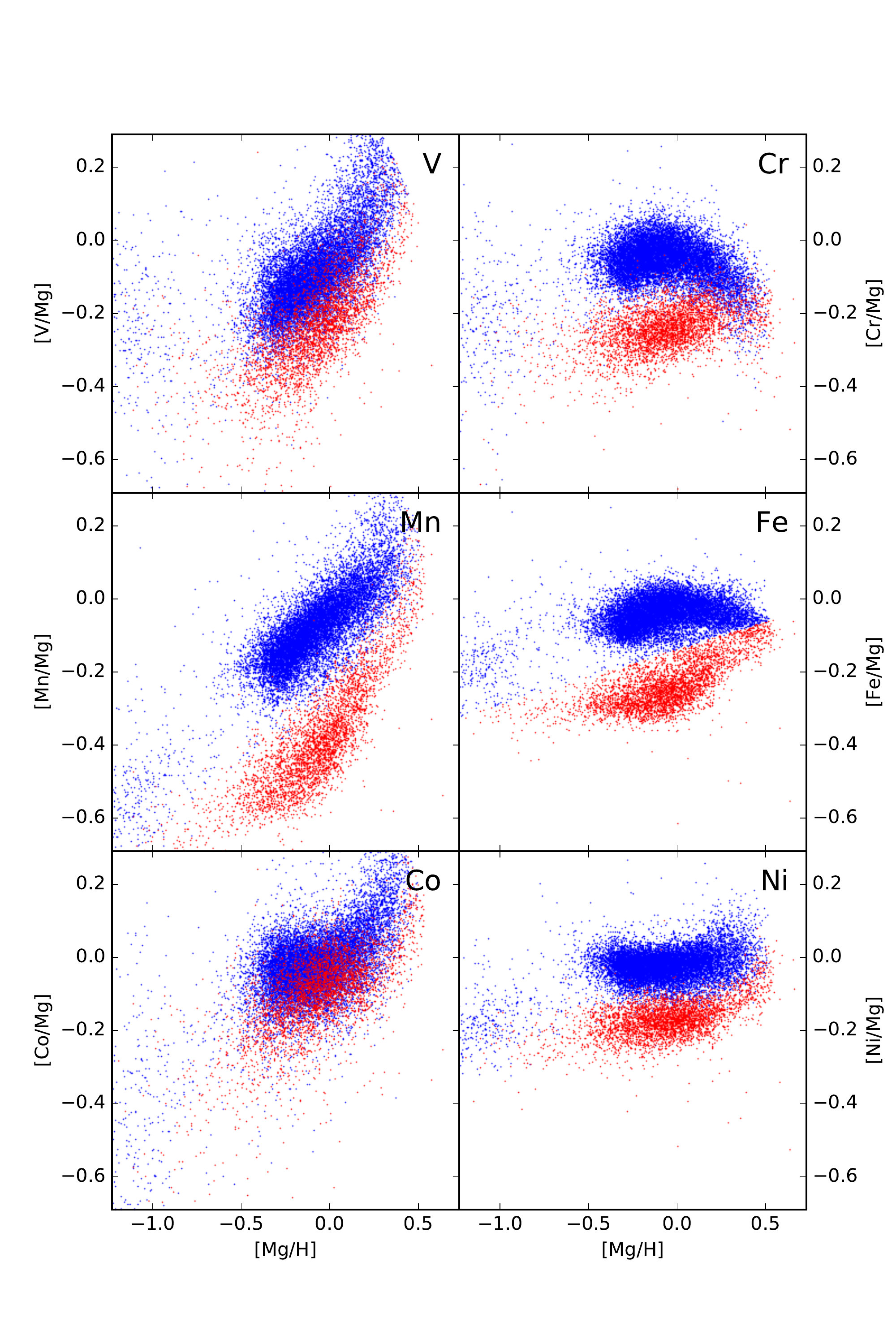} 
\caption{Same as Figure~\ref{fig:alpha_2alpha} for iron-peak elements.
  Elements in the left column have odd atomic numbers; those in the 
  right column have even atomic numbers.
}
\label{fig:fepeak_2alpha} 
\end{figure}

Figure~\ref{fig:fepeak_2alpha} plots \xmg{X} vs. \mgh\ for 
V, Cr, Mn, Fe, Co, and Ni, all ``iron-peak'' elements for which
SNIa are expected to contribute much of the enrichment.
Starting with \femg, we see the familiar two sequences and
the gap between them at sub-solar \mgh, 
though the perfect separation of red and blue points is a
consequence of our defining the boundary between the two populations
in this plane.  The plateau of the low-\femg\ sequence is at
$\femg \approx -0.3$.  The median trend of the high-\femg\
sequence is almost flat, though here as in other studies
(e.g., \citealt{Casagrande2011}) we find a ``banana-shaped''
boundary in which the highest \femg\ stars are concentrated
at solar or slightly sub-solar \mgh, and the upper envelope of the 
population falls slightly at higher and lower \mgh.
Trends for Ni are similar to those for Fe, but the separation
of the two populations is smaller, suggesting a larger relative
CCSN contribution for Ni (see further discussion in \S\ref{sec:twoprocess}).

V, Mn, and Co have odd atomic numbers, and
all three of these elements show increasing \xmg{X} with 
increasing \mgh.  The metallicity-dependence is strongest and
the separation of the two populations largest for Mn, while Co
has the weakest dependence and smallest population separation.
It is not obvious that classic CCSN nucleosynthesis models 
predict a metallicity dependence for odd-$Z$ iron group
elements (see \S VIII.D of \citealt{Woosley2002}).
Cr is an even-$Z$ element, and the \xmg{Cr} tracks are for the
most part similar to the \femg\ tracks. However, there is a significant
downturn of \xmg{Cr} at super-solar \mgh, suggesting that yields
become metallicity-dependent in this regime.

APOGEE abundances for V, Cr, and Ni are generally consistent 
with independent optical analyses of the same stars \citep{Jonsson2018},
and they exhibit similar \xfe{X}-\feh\ trends
to those in local samples. 
APOGEE performs an internal calibration designed to remove abundance
trends with \teff\ in open clusters.
This \teff\ correction is relatively large for Co,
leaving more room for systematic errors in Co abundances 
induced by metallicity-dependent systematics in \teff.
APOGEE \xh{Co} ratios also have significant scatter and a
weak metallicity trend relative to optical measurements.
For Mn, APOGEE measurements agree with independent analyses using
similar methodology.  
However, applying non-LTE corrections to Mn abundances flattens
the observed trend of \xfe{Mn} with \xh{Fe} by boosting Mn
at lower metallicity \citep{Bergemann2008,Battistini2015}.
APOGEE measurements match the optical measurements without
non-LTE corrections.

\begin{figure} 
\includegraphics[width=0.45 \textwidth]{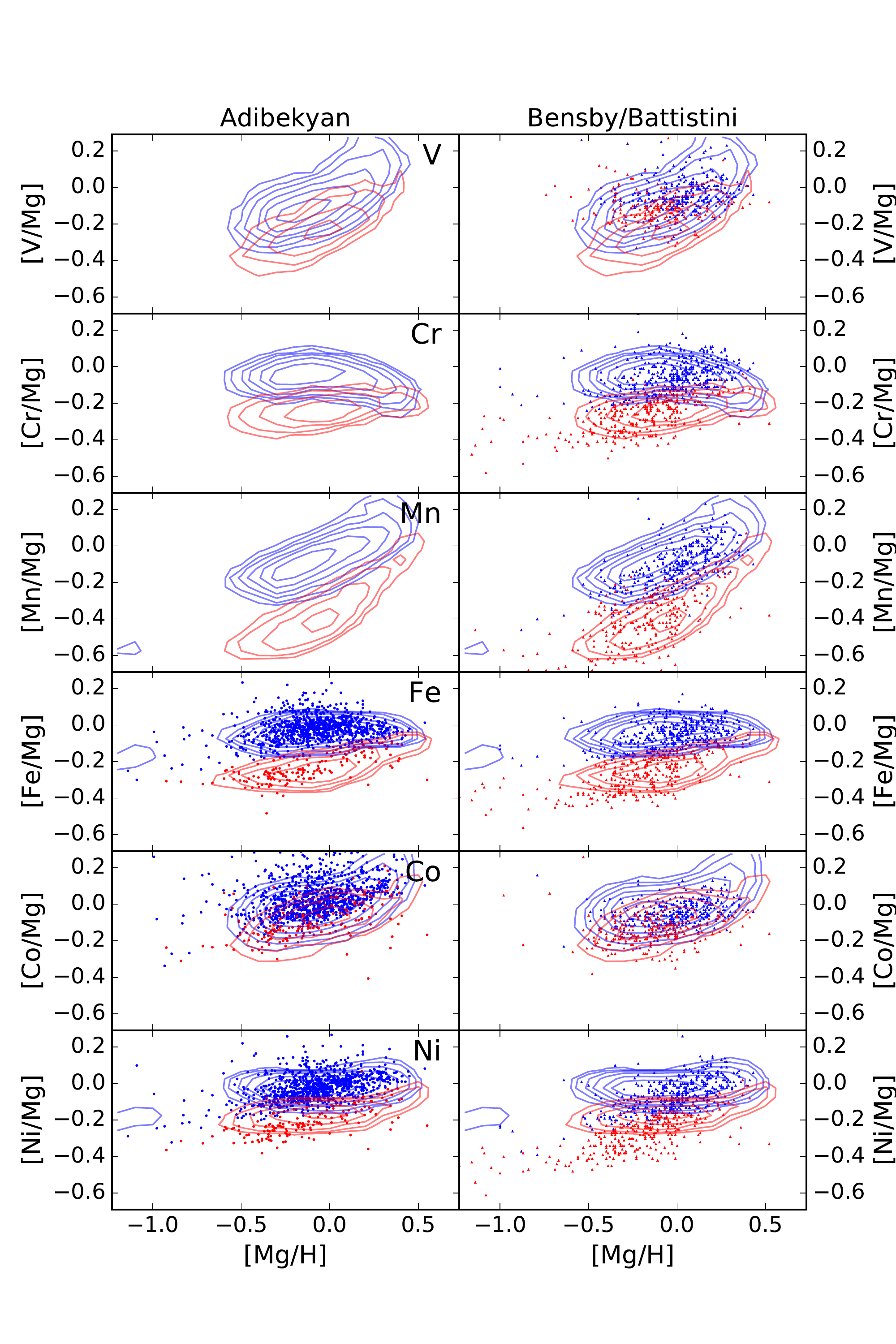} 
\caption{Same as Figure~\ref{fig:alpha_comparison} but for iron-peak elements.
  The \cite{Adibekyan2012} data set does not include V, Cr, or Mn.
  In the right column, results for V, Co, and Mn come from
  \cite{Battistini2015}.
}
\label{fig:fepeak_comparison} 
\end{figure}

Figure~\ref{fig:fepeak_comparison} shows \xmg{X}-\mgh\ 
comparisons to the \cite{Adibekyan2012} and \cite{Bensby2014} samples 
for iron-peak elements.  For V, Mn, and Co, we take values from
\cite{Battistini2015}, who present measurements for a subset of
the \cite{Bensby2014} stars.
Agreement for the iron-peak elements
is generally good, with the most significant difference
being the lowest \xmg{Ni} values in the \cite{Bensby2014} sample.
The separation between the low-\femg\ and high-\femg\ populations
in \xmg{V} and \xmg{Mn} is also smaller for the
\cite{Battistini2015} data than for the APOGEE data.

For the coolest stars in the sample ($T_{\rm eff} < 4000\K$), 
we find that the
\xmg{V}-\mgh\ trends are shifted upward by about 0.1 dex relative
to that of the full sample.  The slopes of these trends and the
separation between the low-\femg\ and high-\femg\ populations
remain similar.  For other iron-peak elements, the trends for
the coolest stars show no obvious offsets or slope differences
relative to the full sample.  The hottest stars ($T_{\rm eff} > 4400\K$)
have slightly higher ($\sim 0.05-0.1$ dex) \xmg{V} at $\mgh \approx -0.3$
but show no other obvious offsets or slope differences.

\section{A semi-empirical model of element yields}
\label{sec:twoprocess}

\subsection{Model definition}

As a semi-empirical description of our observed trends that incorporates 
basic physical expectations about the origin of these elements,
we adopt a ``2-process model'' in which each star's abundances are 
represented as the sum of a ``core collapse process'' 
with amplitude $\Acc$ and an ``SNIa process'' with amplitude $\AIa$,
with these amplitudes multiplying coefficients $p^X$ that are specific
to each element.  The abundance of element $X$ is 
\begin{equation} 
\left({{\rm X} \over {\rm H}}\right) = \Acc \pxcc(Z) + \AIa \pxIa(Z)~. 
\label{eqn:xh} 
\end{equation} 
The dependence of each process $\pxcc$ and $\pxIa$ on $Z$ allows 
for metallicity-dependent IMF-averaged yields of element X.
The model makes no specific assumption about the chemical evolution
history that leads to a given star's value of $\Acc$ and $\AIa$
(see further discussion in \S\ref{sec:twoprocess_metdepyield}).
Our comments about theoretical expectations throughout this section
are again based on the references discussed at the beginning
of \S\ref{sec:trends}.

We normalize $\Acc=1$ at solar abundance, so
\begin{equation}
\pxccs \equiv \pxcc(\Zsun) = \left({{\rm X}\over{\rm H}}\right)_{{\rm cc},\odot}
\label{eqn:pxccs}
\end{equation}
is the contribution of core collapse supernovae to the solar abundance
ratio of element X.
With this normalization, one can take $({\rm X}/{\rm H})$ to 
represent either a number density ratio or a mass-fraction ratio.
We assume (1) that Mg is a pure CCSN element with an IMF-averaged yield
that is independent of metallicity, (2) that the IMF-averaged CCSN and
SNIa iron yields are independent of metallicity, and (3) that the
observed high-$\alpha$ plateau at $\mgfe=+0.3$ corresponds to pure
CCSN enrichment.  One can therefore infer $\Acc$ and $\AIa$ for a given
star from its Mg abundance and its \mgfe\ ratio:
\begin{equation}
\Acc = 10^{\mgh}~,
\label{eqn:Acc}
\end{equation}
\begin{equation} 
{\Acc \over \Acc + \AIa} = 10^{\mgfe - 0.3}~,
\label{eqn:snratio} 
\end{equation} 
or equivalently
\begin{equation}
  {\AIa\over\Acc} = 10^{0.3-\mgfe}-1~.
  \label{eqn:aratio}
\end{equation}
Since $10^{0.3} \approx 2$, Equation~(\ref{eqn:aratio}) implies
that $\AIa \approx \Acc = 1$ for a star with solar abundances.

In principle one can generalize this model to allow metallicity-dependent
Mg or Fe yields or a different plateau value of \mgfe,
but we will not do so here.  For other elements we do allow metallicity
dependence, which for simplicity we model as power laws of slopes
$\alphacc$ and $\alphaIa$ between abundance and (Mg/H):
\begin{eqnarray} 
\pxcc(Z) &=& \pxccs \cdot 10^{\alphacc\mgh} 
\label{eqn:alphaxccdef} \\
\pxIa(Z) &=& \pxIas \cdot 10^{\alphaIa\mgh} ~.
\label{eqn:alphaxIadef}
\end{eqnarray} 
For each element, there are three model parameters: the slopes
$\alphacc$ and $\alphaIa$ and the relative SNIa contribution
at solar metallicity, described by the ratio
\begin{equation} 
\Rxsun \equiv \frac{\pxIas}{\pxccs}~. 
\label{eqn:Rxdef} 
\end{equation} 
The normalization convention of Equation~(\ref{eqn:pxccs}) together with
$A_{{\rm Ia},\odot}=A_{{\rm cc},\odot}=1$ implies
\begin{equation}
\pxccs + \pxIas = \left({{\rm X}\over{\rm H}}\right)_\odot~.
\label{eqn:pnorm}
\end{equation}

The ratio $\AIa/\Acc$ (eq.~\ref{eqn:aratio}) describes the relative
contribution of the SNIa and CCSN processes 
{\it to a given star}, while $\Rxsun$ describes the
relative contribution of these two processes {\it to a given element} X
at solar \mgh.
Our model assumptions (1) and (2) correspond to parameter values
$R_{\rm Ia}^{{\rm Mg}}= \alphacc^{{\rm Mg}}=\alphaIa^{{\rm Mg}}=0$ and
$\alphacc^{{\rm Fe}} = \alphaIa^{{\rm Fe}} = 0$, respectively,
while assumption (3) implies $R_{\rm Ia}^{{\rm Fe}} = 1$.
The metallicity dependence of an $\xmg{X}-\mgh$ trend is related to but not
the same as the implied metallicity dependence of the yield
of element $X$, a point we discuss further in 
\S\ref{sec:twoprocess_metdepyield}.

Once the model parameters are determined, one can use a star's
measured \mgh\ and \mgfe\ to predict the abundances of other elements.
Specifically, because $p^{\rm Mg}_{\rm Ia} = \alphacc^{{\rm Mg}} = 0$,
one has a relation for the ratio
\begin{equation} 
\left({X \over {\rm Mg}}\right) = 
       {\Acc\pxcc(Z)+\AIa\pxIa(Z) \over 
       \Acc p^{\rm Mg}_{\rm cc} } ~.
\label{eqn:xmgratio} 
\end{equation} 
Scaling to solar abundances and using $(\Acc/\AIa)_\odot = 1$ gives
\begin{align}
{({\rm X}/{\rm Mg}) \over ({\rm X}/{\rm Mg})_\odot} &=
  {\pxcc(Z)+(\AIa/\Acc)\pxIa(Z) \over \pxcc(\Zsun) + \pxIa(\Zsun)} \\ &=
  {\pxcc(Z) \over \pxcc(\Zsun)} \times {1+(\AIa/\Acc)\pxIa(Z)/\pxcc(Z) \over
  1+\pxIa(\Zsun)/\pxcc(\Zsun)}~.  
\end{align}
One can combine this result with the 
definitions~(\ref{eqn:alphaxccdef})-(\ref{eqn:Rxdef}) to obtain
\begin{align}
[{\rm X}/{\rm Mg}] &= \alpha_{\rm cc}\mgh \nonumber \\
  &+ \log\left[{1 + \Rxsun (\AIa/\Acc) \cdot
	       10^{(\alpha_{\rm Ia}-\alpha_{\rm cc})\mgh} \over 1+\Rxsun }
	       \right]~, 
\label{eqn:xmg} 
\end{align}
where the value of $\AIa/\Acc$ is inferred from the \mgfe\ ratio
via Equation~(\ref{eqn:aratio}).  

For a pure CCSN element, with $\Rxsun = 0$, Equation~(\ref{eqn:xmg})
implies $\xmg{X} = \alphacc\mgh$.  If the metallicity dependence
is also $\alphacc=0$, the element simply tracks Mg,
with $\xmg{X}=0$ for all stars regardless of their SNIa enrichment.
For any element with metallicity-independent yields, Equation~(\ref{eqn:xmg})
simplifies to
\begin{equation}
[{\rm X}/{\rm Mg}] = \log\left[{{1 + \Rxsun (\AIa/\Acc)} \over 
                               {1+\Rxsun}} \right]~,
\label{eqn:xmgnomet} 
\end{equation}
which leads to $\xmg{X} = -\log(1+\Rxsun)$ for a ``plateau'' star with
no SNIa enrichment and $\xmg{X}=0$ for any star with
equal SNIa and CCSN amplitudes ($\AIa/\Acc=1$).
For example, Fe has $\Rxsun=1$, and Equation~(\ref{eqn:xmgnomet})
implies $\femg=-0.3$ for pure CCSN enrichment and $\femg=0$
for equal SNIa and CCSN contributions.

\subsection{$\alpha$-elements}
\label{sec:twoprocess_alpha}

The parameters of the 2-process model, different for each element, 
can in principle be fit to the entire sample of stars, without
dividing them into low-\femg\ and high-\femg\ populations.  Here,
however, we fit the model to the median sequences of the two 
populations, i.e., to the median \xmg{X} values of ${\rm SNR}>200$
stars in 0.1-dex bins of \mgh\ over the range $-0.8 \leq \mgh \leq +0.5$.
The median sequences themselves are listed in 
Tables~\ref{tbl:medseq_alpha_lowFe}-\ref{tbl:medseq_peak_highFe}.
We perform unweighted least-squares fits; since the model is simple
and the (tiny) statistical errors on the median abundance ratios
are smaller than systematic errors, a weighted $\chi^2$ fit is
unwarranted.  All of the model fits are poor in a formal $\chi^2$ sense.
We perform one set of fits with the SNIa metallicity index
set to $\alpha_{\rm Ia}=0$ so that there are only two parameters,
and a second set of fits with free $\alpha_{\rm Ia}$.
We find best-fit parameters by a simple grid search with steps of
0.01 in each free parameter, and we do not infer parameter errors
because the fits are formally poor and we do not expect the
model to be a complete description of the data.

\begin{table}
\centering
\caption{Median sequences, $\alpha$-elements, low-\femg\ stars}
\begin{tabular}{rrrrr}
\hline
\hline
[Mg/H]  & [O/Mg] & [Si/Mg] & [S/Mg] & [Ca/Mg] \\
\hline
-0.744  &  0.037 &  0.061 &  0.074 & -0.108 \\
-0.639  &  0.010 &  0.000 &  0.043 & -0.130 \\
-0.541  & -0.015 & -0.021 &  0.051 & -0.137 \\
-0.437  & -0.017 & -0.032 &  0.072 & -0.128 \\
-0.349  & -0.019 & -0.034 &  0.020 & -0.132 \\
-0.254  & -0.024 & -0.047 &  0.027 & -0.132 \\
-0.146  & -0.028 & -0.065 &  0.006 & -0.142 \\
-0.050  & -0.026 & -0.072 & -0.032 & -0.142 \\
 0.044  & -0.029 & -0.082 & -0.046 & -0.144 \\
 0.146  & -0.028 & -0.090 & -0.073 & -0.138 \\
 0.237  & -0.030 & -0.096 & -0.089 & -0.125 \\
 0.346  & -0.043 & -0.102 & -0.108 & -0.115 \\
 0.448  & -0.030 & -0.095 & -0.099 & -0.100 \\
\hline
\end{tabular}
\tablecomments{Median \xmg{X} ratios of $\alpha$-elements
for stars in the low-\femg\ (high-$\alpha$) population with
SNR$>200$, computed in 0.1-dex bins of \mgh.  The SNR
threshold is lowered to 100 for the first two bins,
with $\mgh < -0.6$.  Median values
of \mgh\ within each bin are given in the left column.
}
\label{tbl:medseq_alpha_lowFe}
\end{table}

\begin{table}
\centering
\caption{Median sequences, $\alpha$-elements, high-\femg\ stars}
\begin{tabular}{rrrrr}
\hline
\hline
[Mg/H]  & [O/Mg] & [Si/Mg] & [S/Mg] & [Ca/Mg] \\
\hline
-0.744  &  0.064 &  0.051 &  0.134 & -0.042 \\
-0.639  &  0.036 &  0.023 &  0.089 & -0.057 \\
-0.541  &  0.026 &  0.037 &  0.051 & -0.060 \\
-0.437  &  0.023 &  0.029 &  0.036 & -0.055 \\
-0.349  &  0.019 &  0.018 &  0.030 & -0.048 \\
-0.254  &  0.018 &  0.011 &  0.015 & -0.037 \\
-0.146  &  0.017 &  0.006 &  0.002 & -0.022 \\
-0.050  &  0.015 &  0.000 & -0.008 & -0.014 \\
 0.044  &  0.009 & -0.011 & -0.024 & -0.015 \\
 0.146  &  0.002 & -0.022 & -0.035 & -0.018 \\
 0.237  & -0.000 & -0.029 & -0.041 & -0.017 \\
 0.346  & -0.011 & -0.037 & -0.051 & -0.015 \\
 0.448  & -0.024 & -0.058 & -0.067 & -0.033 \\
\hline
\end{tabular}
\tablecomments{Same as Table~\ref{tbl:medseq_alpha_lowFe} but for stars
in the high-\femg\ (low-$\alpha$) population.
}
\label{tbl:medseq_alpha_highFe}
\end{table}

\begin{figure} 
\includegraphics[width=0.45 \textwidth]{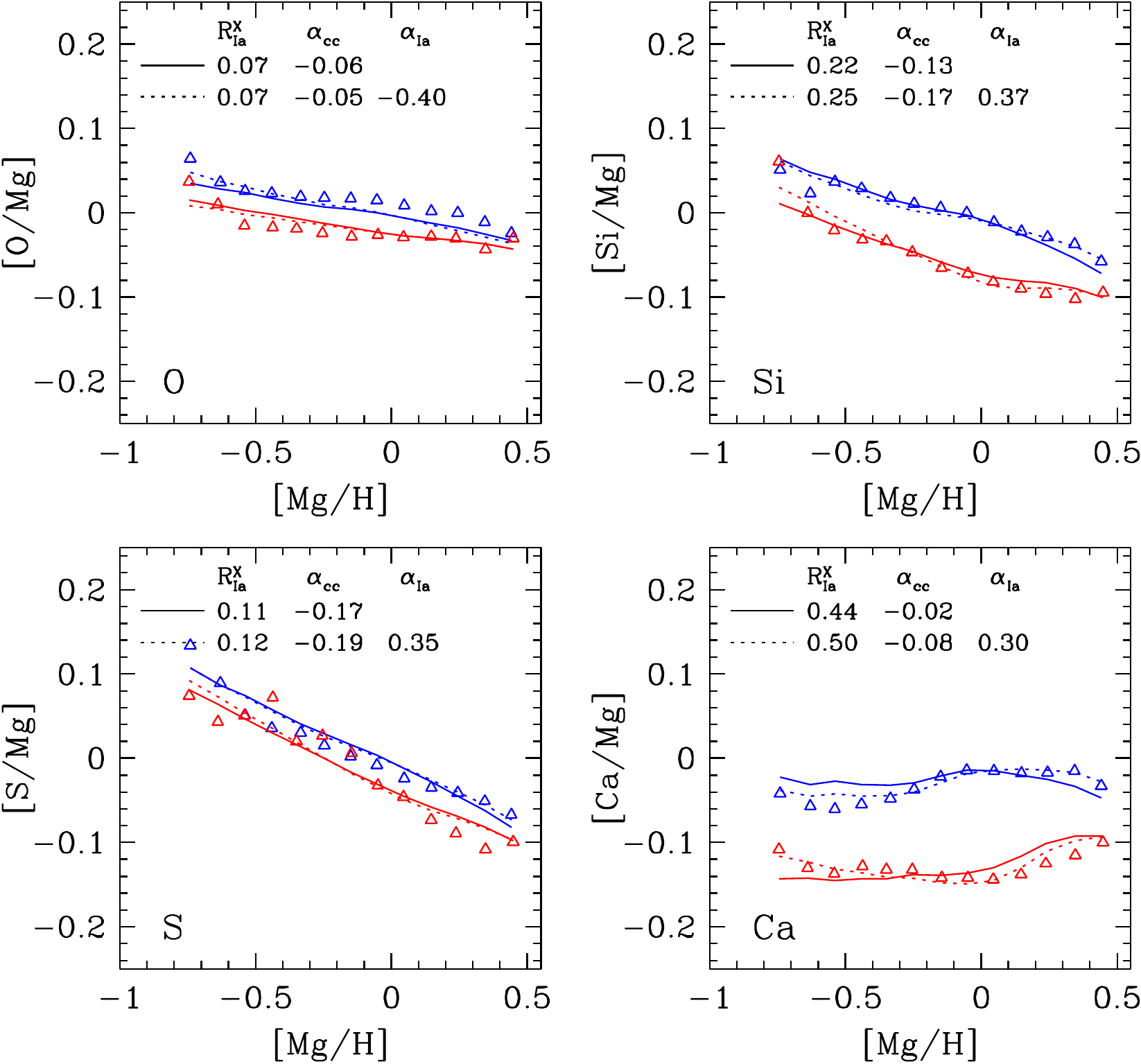}
\caption{Median sequences and the 2-process model for $\alpha$-elements.
Triangles show the median \xmg{X} of stars in 0.1-dex bins of \mgh\ in
the low-\femg\ (red) and high-\femg\ (blue) populations, respectively.
These sequences are derived from the same set of stars plotted 
in Figure~\ref{fig:alpha_2alpha}.  Solid curves show fits of the
2-process model to the median sequences with the metallicity-dependence
of the SNIa contribution set to zero.
Dotted curves show fits in which the index $\alpha_{\rm Ia}$ of
the SNIa metallicity dependence is free.  The fitted values of
$\Rxsun$, the ratio of amplitudes of the SNIa process to the CCSN process
at $\mgh=0$, are listed in each panel, together with the fitted
metallicity slopes $\alpha_{\rm cc}$ and $\alpha_{\rm Ia}$
(see eqs.~\ref{eqn:alphaxccdef}-\ref{eqn:Rxdef}).
Of these elements, Ca has the largest relative SNIa contribution,
and Si and S have the strongest metallicity dependence.
} 
\label{fig:twoprocess_alpha} 
\end{figure}

Figure~\ref{fig:twoprocess_alpha} shows results for the $\alpha$-elements,
for which the full $\xmg{X}-\mgh$ distributions were shown
previously in Figure~\ref{fig:alpha_2alpha}.
The small separation of the two \xmg{O} median sequences leads
to a small but non-zero value of $\Rxsun=0.07$.  Even this level
of SNIa contribution is higher than theoretically expected.
The sequence separation and inferred $\Rxsun$ could represent
a contribution to O from some other source, such as AGB stars,
that has a delayed enrichment time profile resembling that
of SNIa.  Alternatively, the separation could arise from a systematic error
in our oxygen abundance determinations that is correlated
with \femg, such that we systematically underestimate \xmg{O}
in low-\femg\ stars, though we have not identified a systematic that
would lead to this behavior
(see previous discussion in \S\ref{sec:trends_alpha}).
The median trends show a small but clear
metallicity dependence at $\mgh< -0.3$, but this dependence is
not well described by a power law over the full \mgh\ range
so our model does not fit it well.

For Si our fit implies a larger but still sub-dominant SNIa contribution,
$\Rxsun = 0.22-0.25$, qualitatively consistent with expectations.
There is a significant trend of decreasing \xmg{Si} with increasing
\mgh.  The fit with free $\alpha_{\rm Ia}$ better describes
the full locus of the high-\femg\ sequence, but only at the 
0.04-dex level, and given the uncertainties of the data and the
simplicity of the power-law model it is not clear that this
improvement is physically significant.  The decreasing yield
for CCSN appears more robust, and the inferred slope $\alpha_{\rm cc}$ is
similar whether or not $\alpha_{\rm Ia}$ is forced to zero
($-0.13$ vs. $-0.17$).  We emphasize again that our model
assumes a metallicity-independent yield for Mg by construction,
and one should more accurately view the value of $\alpha_{\rm cc}$
as reflecting the metallicity dependence relative to that of Mg.

As previously noted, the similar \xmg{S} of the two populations
implies, at face value, very little SNIa contribution to sulfur
enrichment; our model fits yield $\Rxsun=0.11$ or 0.12.
This result runs somewhat counter to theoretical yield models, which
predict that the relative SNIa contribution increases,
or at least does not decrease, for heavier $\alpha$-elements.
The inferred metallicity dependence,
$\alpha_{\rm cc} = -0.17$ or $-0.19$, is similar to that for Si.
For Ca we infer a larger $\Rxsun = 0.44-0.50$, as expected for
this heavier element, and weaker metallicity dependence.
In this case, allowing free $\alpha_{\rm Ia}$ produces a 
qualitatively better fit for both sequences across the
full \mgh\ range, with inferred $\alpha_{\rm Ia}=0.30$.

O and Mg are produced primarily during the hydrostatic evolution
of massive stars, before they explode as CCSN, and their expected yields
increase rapidly with progenitor mass. Si and Ca, on
the other hand, are produced mainly by explosive nucleosynthesis
during the CCSN event itself, and their expected yields are less 
mass-dependent.  The hydrostatic/explosive element ratio can
therefore provide a diagnostic of the stellar IMF; for example,
\cite{McWilliam2013}, \cite{Vincenzo2015}, and \cite{Carlin2018}
have used this ratio to argue that the IMF in the Sagittarius
dwarf galaxy was deficient in high mass stars relative to the
Milky Way disk.  In principle, the offsets found here for Si and Ca
could arise from a change in the IMF between the low-\femg\ and
high-\femg\ populations.  However, given the constancy of these
trends through the disk and the clear dependence on \femg, it
is more natural to associate them with SNIa contributions to
these elements, which are theoretically expected in any case.
These SNIa contributions should be accounted for when using
the hydrostatic/explosive ratio to test for IMF variations,
and our 2-process model fits provide a convenient way to do so.

\begin{table}
\centering
\caption{Median sequences, odd-$Z$ elements, low-\femg\ stars}
\begin{tabular}{rrrrr}
\hline
\hline
[Mg/H]  & [Na/Mg] & [Al/Mg] & [P/Mg] & [K/Mg] \\
\hline
-0.744  & ....... & -0.156 & -0.458 & -0.164 \\
-0.639  & ....... & -0.108 & -0.303 & -0.120 \\
-0.541  & -0.388 & -0.122 & -0.333 & -0.118 \\
-0.437  & -0.344 & -0.111 & -0.285 & -0.104 \\
-0.349  & -0.329 & -0.101 & -0.281 & -0.101 \\
-0.254  & -0.301 & -0.065 & -0.266 & -0.087 \\
-0.146  & -0.280 & -0.059 & -0.284 & -0.087 \\
-0.050  & -0.226 & -0.029 & -0.280 & -0.077 \\
 0.044  & -0.230 &  0.008 & -0.274 & -0.066 \\
 0.146  & -0.297 &  0.024 & -0.262 & -0.071 \\
 0.237  & -0.161 &  0.029 & -0.263 & -0.062 \\
 0.346  & -0.110 &  0.055 & -0.279 & -0.063 \\
 0.448  &  0.010 & ...... & -0.387 & -0.024 \\
\hline
\end{tabular}
\tablecomments{Same as Table~\ref{tbl:medseq_alpha_lowFe}
but for light odd-$Z$ elements.  Blank entries 
mark bins where the median value is a non-detection.
}
\label{tbl:medseq_oddz_lowFe}
\end{table}

\begin{table}
\centering
\caption{Median sequences, odd-$Z$ elements, high-\femg\ stars}
\begin{tabular}{rrrrr}
\hline
\hline
[Mg/H]  & [Na/Mg] & [Al/Mg] & [P/Mg] & [K/Mg] \\
-0.744  & -0.143 & -0.269 &  0.068 & -0.082 \\
-0.639  & -0.174 & -0.149 & -0.013 & -0.041 \\
-0.541  & -0.137 & -0.077 &  0.019 & -0.047 \\
-0.437  & -0.112 & -0.081 & -0.022 & -0.054 \\
-0.349  & -0.106 & -0.059 & -0.036 & -0.044 \\
-0.254  & -0.099 & -0.052 & -0.040 & -0.038 \\
-0.146  & -0.091 & -0.042 & -0.038 & -0.030 \\
-0.050  & -0.099 & -0.025 & -0.044 & -0.027 \\
 0.044  & -0.107 &  0.001 & -0.070 & -0.028 \\
 0.146  &  0.011 &  0.024 & -0.111 & -0.025 \\
 0.237  &  0.069 &  0.049 & -0.148 & -0.011 \\
 0.346  &  0.067 &  0.069 & -0.200 &  0.001 \\
 0.448  & -0.017 & ...... & -0.256 &  0.014 \\
\hline
\end{tabular}
\tablecomments{Same as Table~\ref{tbl:medseq_oddz_lowFe} but for stars
in the high-\femg\ (low-$\alpha$) population.
}
\label{tbl:medseq_oddz_highFe}
\end{table}

\begin{figure} 
\includegraphics[width=0.45 \textwidth]{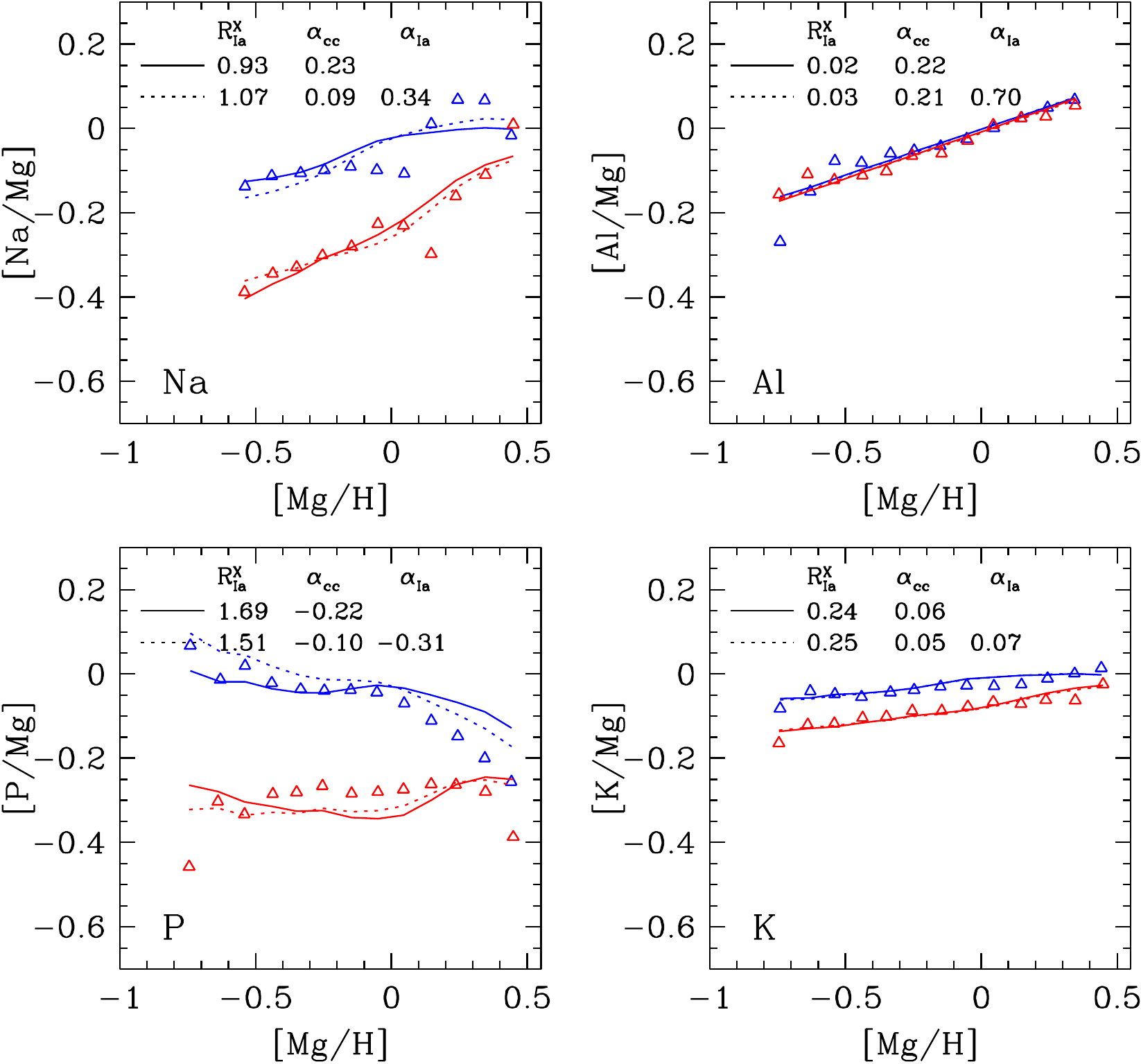}
\caption{Same as Figure~\ref{fig:twoprocess_alpha} but for 
light odd-$Z$ elements.
The unusual feature in the observed median trend for \xmg{Na} is
caused by the analysis artifact discussed in \S\ref{sec:trends_oddz}.
Na and P exhibit a large sequence separation implying a large
SNIa contribution, contrary to theoretical expectations.
Al and K have small inferred SNIa contributions and slowly
rising metallicity trends.
} 
\label{fig:twoprocess_oddz} 
\end{figure}

\subsection{Light odd-$Z$ elements}
\label{sec:twoprocess_oddz}

Figure~\ref{fig:twoprocess_oddz} presents similar median sequences and
model fits for the light odd-$Z$ elements.  Note that the \xmg{X}
axis-range is expanded on these plots relative to the $\alpha$-element plots.
For sodium the large
gap in \xmg{Na} between the low-\femg\ and high-\femg\ sequences
implies, in the context of our model, a large SNIa contribution,
with inferred $\Rxsun = 0.93-1.07$.  These values imply that
slightly over half of the sodium in solar metallicity stars 
would come from SNIa, in strong disagreement with supernova models.
Some Na production from AGB stars is expected \citep{Ventura2005,Karakas2010},
but at least in the calculations of \cite{Andrews2017} and
\cite{Rybizki2017} these predicted contributions are much smaller
than the CCSN contribution.
The broad trend of increasing \xmg{Na} with increasing \mgh\ 
is expected for an odd-$Z$ element.  The sharp changes of slope
near $\mgh=0$ on both sequences are affected by the previously discussed
abundance artifact evident in Figure~\ref{fig:oddz_2alpha}.
Unsurprisingly, the power-law metallicity dependence of the
model is a poor fit to these median sequences, and the values
of $\alphacc$ and $\alphaIa$ should be considered unreliable.
Statistical errors for the Na abundances are large (mean of 0.068 dex)
because the Na lines in the APOGEE spectral range are weak.
However, there is no obvious systematic error that would cause
a difference of \xmg{Na} for stars in the low-\femg\ and high-\femg\
populations, so the implication of a substantial non-CCSN 
contribution to sodium enrichment appears more robust.

For aluminum we find a low SNIa contribution, with best fit
$\Rxsun=0.02$, and an increasing metallicity trend with
$\alpha_{\rm cc}=0.22$.  The temperature effects
discussed in \S\ref{sec:trends_oddz} imply some systematic uncertainty in
$\alpha_{\rm cc}$, as the slope for the coolest stars in our sample
is steeper than that of the full sample.
For potassium we find a larger
but sub-dominant SNIa contribution, $\Rxsun=0.24$, and a
positive but weak metallicity trend, with $\alpha_{\rm cc}=0.06$.
Existing supernova yield calculations underpredict observed potassium
abundances by a substantial factor
(see, e.g., fig.~9 of \citealt{Andrews2017} or fig.~14
of \citealt{Rybizki2017}).

For phosphorous, as for sodium, the large gap between the
two median sequences implies a large SNIa contribution in
the context of our model, with $\Rxsun \approx 1.6$.
This is again surprising relative to model predictions
\citep{Rybizki2017}, in which CCSN contributions dominate
over SNIa and AGB production.
The metallicity dependence of the high-\femg\ sequence is
curved and not well fit by our power-law model, even with
$\alpha_{\rm cc}$ and $\alpha_{\rm Ia}$ both free to vary.
As with sodium, statistical errors (mean of 0.076 dex)
and potential systematic errors are large, so we are cautious
about drawing physical conclusions from the phosphorous results.
However, there is no obvious systematic that would lead
to different \xmg{P} ratios for the low-\femg\ and high-\femg\ 
populations.

\begin{table*}
\centering
\caption{Median sequences, iron-peak elements, low-\femg\ stars}
\begin{tabular}{rrrrrrr}
\hline
\hline
[Mg/H]  & [V/Mg] & [Cr/Mg] & [Mn/Mg] & [Fe/Mg] & [Co/Mg] & [Ni/Mg] \\
\hline
-0.744  & -0.403 & -0.277 & -0.615 & -0.298 & -0.389 & -0.260 \\
-0.639  & -0.374 & -0.292 & -0.566 & -0.292 & -0.282 & -0.215 \\
-0.541  & -0.363 & -0.303 & -0.536 & -0.295 & -0.244 & -0.203 \\
-0.437  & -0.355 & -0.261 & -0.523 & -0.285 & -0.194 & -0.191 \\
-0.349  & -0.336 & -0.268 & -0.499 & -0.281 & -0.177 & -0.182 \\
-0.254  & -0.307 & -0.257 & -0.468 & -0.266 & -0.136 & -0.174 \\
-0.146  & -0.273 & -0.249 & -0.443 & -0.263 & -0.096 & -0.175 \\
-0.050  & -0.225 & -0.240 & -0.402 & -0.253 & -0.072 & -0.171 \\
 0.044  & -0.193 & -0.231 & -0.351 & -0.236 & -0.058 & -0.168 \\
 0.146  & -0.155 & -0.215 & -0.273 & -0.203 & -0.052 & -0.161 \\
 0.237  & -0.116 & -0.209 & -0.206 & -0.171 & -0.045 & -0.146 \\
 0.346  & -0.044 & -0.205 & -0.143 & -0.151 & -0.020 & -0.132 \\
 0.448  & ...... & -0.233 & -0.074 & -0.146 &  0.066 & -0.129 \\
\hline
\end{tabular}
\tablecomments{Same as Table~\ref{tbl:medseq_alpha_lowFe}
but for iron-peak elements.  Blank entries
mark bins in which the median value is a non-detection.
}
\label{tbl:medseq_peak_lowFe}
\end{table*}

\begin{table*}
\centering
\caption{Median sequences, iron-peak elements, high-\femg\ stars}
\begin{tabular}{rrrrrrr}
\hline
\hline
[Mg/H]  & [V/Mg] & [Cr/Mg] & [Mn/Mg] & [Fe/Mg] & [Co/Mg] & [Ni/Mg] \\
\hline
-0.744  & -0.163 & -0.123 & -0.224 & -0.054 & -0.221 & -0.035 \\
-0.639  & -0.194 & -0.072 & -0.209 & -0.067 & -0.167 & -0.014 \\
-0.541  & -0.222 & -0.044 & -0.178 & -0.058 & -0.090 &  0.002 \\
-0.437  & -0.204 & -0.058 & -0.176 & -0.062 & -0.071 & -0.011 \\
-0.349  & -0.175 & -0.059 & -0.164 & -0.061 & -0.047 & -0.018 \\
-0.254  & -0.149 & -0.049 & -0.140 & -0.053 & -0.037 & -0.024 \\
-0.146  & -0.125 & -0.036 & -0.102 & -0.037 & -0.036 & -0.027 \\
-0.050  & -0.090 & -0.033 & -0.063 & -0.023 & -0.030 & -0.026 \\
 0.044  & -0.051 & -0.043 & -0.025 & -0.022 & -0.004 & -0.019 \\
 0.146  &  0.007 & -0.065 &  0.010 & -0.028 &  0.031 & -0.009 \\
 0.237  &  0.047 & -0.096 &  0.047 & -0.035 &  0.076 & -0.004 \\
 0.346  & -0.034 & -0.129 &  0.082 & -0.046 &  0.125 &  0.001 \\
 0.448  & ...... & -0.165 &  0.057 & -0.066 &  0.111 & -0.018 \\
\hline
\end{tabular}
\tablecomments{Same as Table~\ref{tbl:medseq_peak_lowFe}
but for high-\femg\ stars.
}
\label{tbl:medseq_peak_highFe}
\end{table*}

\begin{figure}
\includegraphics[width=0.45 \textwidth]{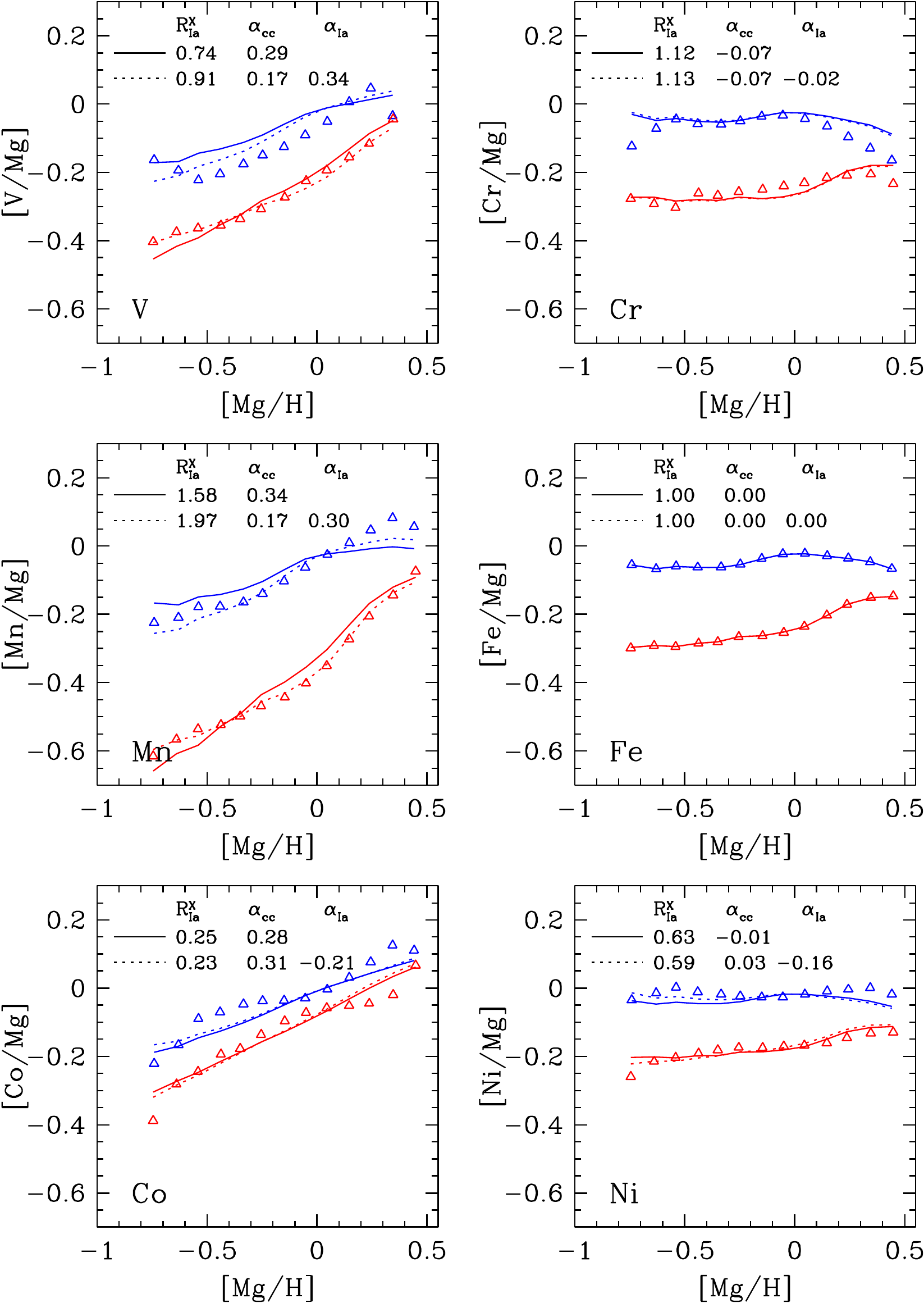} 
\caption{Same as Figure~\ref{fig:twoprocess_alpha} but for iron-peak elements.
  With the exception of Co, all elements show substantial SNIa contributions,
  with $\Rxsun > 0.5$.  The odd-$Z$ elements (left column) show strong
  metallicity trends while the even-$Z$ elements (right column) do not.
}
\label{fig:twoprocess_peak} 
\end{figure}

\subsection{Iron-peak elements}
\label{sec:twoprocess_fepeak}

Figure~\ref{fig:twoprocess_peak} shows median sequences for iron-peak
elements, with the odd-$Z$ elements V, Mn, Co, on the left and
the even-$Z$ elements Cr, Fe, Ni on the right.  
By construction the model for Fe has $\Rxsun=1.0$,
$\alphacc=\alphaIa=0.0$, and fits the observed median sequences perfectly.
For Cr the inferred process ratio is $\Rxsun=1.12$; the metallicity trend is
mostly flat, but there is a downturn of \xmg{Cr} at super-solar
\mgh\ in the high-\femg\ population, similar to the behavior for P.
For Ni the inferred $\Rxsun=0.63$, implying a larger CCSN 
contribution.  The inferred metallicity dependence is again
weak.  Note that the median sequence for the low-\femg\ stars
(red points) turns up at high \mgh\ even with $\alphacc=\alphaIa=0$ 
because the value of \femg\ is higher, i.e., the upturn reflects
the larger contribution of SNIa elements at the high-metallicity
end of the low-\femg\ sequence.

All of the odd-$Z$ iron-peak elements show clear metallicity trends,
steeper than those of the light odd-$Z$ elements.  The large
separation of the median sequences for Mn implies a large 
SNIa contribution, larger than that of any other element
considered here, with $\Rxsun=1.58$ (1.97) for the fit with
$\alphaIa=0$ ($\alphaIa$ free). 
The inferred $\Rxsun$ for V and Co are 0.74 and 0.25,
respectively (for $\alphaIa=0$).  In the case of Mn, allowing a metallicity
trend for the SNIa process noticeably improves the fit to the 
observations.  For V and Co the improvement is less clear, though
the inference of a metallicity-dependent CCSN process is robust.

\cite{Battistini2015} have examined \xfe{X} trends for V, Mn, and Co
in the local Galactic disk, in comparison to bulge, halo, and
globular cluster samples.  For Mn they find a significant rise
of \xfe{Mn} with increasing \feh, qualitatively similar to our
results for \xmg{Mn}.  However, the trend largely disappears
if they include non-LTE corrections in their Mn abundance 
determinations.  Their trends for V and Co are closer to those of 
$\alpha$-elements, with decreasing \xfe{X} over the range
$-0.8 > \feh > 0$.  In light of our results, it seems that
the different behavior of V and Co relative to Mn might be
explained by the combination of a lower SNIa fraction and
a weaker metallicity dependence.  This combination allows an underlying
positive metallicity trend to be masked by the trend of
higher SNIa iron contribution at higher \feh, analogous
to our results for \xfe{Al} shown in Figure~\ref{fig:AlFeMg}.
However, all three of these element abundances are challenging
to measure, so further investigation using both optical and
near-IR samples is warranted.

\subsection{Relation of median sequences to metallicity-dependent yields}
\label{sec:twoprocess_metdepyield}

Figs.~\ref{fig:twoprocess_alpha}-\ref{fig:twoprocess_peak} suggest
metallicity-dependent yields for some elements, especially the
odd-$Z$ elements below and near the iron peak.
However, the trend of an element ratio with $\mgh$ depends on the
chemical evolution history as well as on the yield metallicity-dependence
itself, because the metals present in the interstellar medium (ISM)
at a given time come
from stars that formed over a range of previous times, when the
composition of the ISM (and thus of newly forming stars) may have
been different.  Thus, even if the stellar IMF and nucleosynthetic 
processes within stars remain constant throughout the Galactic disk, 
one might expect location-dependent trends of \xmg{X} with \mgh\ 
because of differences in star formation history, accretion, or outflows.  
To address this issue, we use one-zone chemical evolution models similar to
those of \cite{Weinberg2017}; we compute their evolution
numerically instead of analytically so that we can include
a hypothetical element with a metal-dependent yield.
In these models, the predicted \xmg{X}-\mgh\ trends prove
insensitive to details of the evolutionary history, which makes
the relation of observed trends to IMF-averaged yields straightforward
to interpret and largely explains why we find median sequences
that are constant throughout the Galactic disk.

\begin{figure}
  \includegraphics[width=0.45 \textwidth]{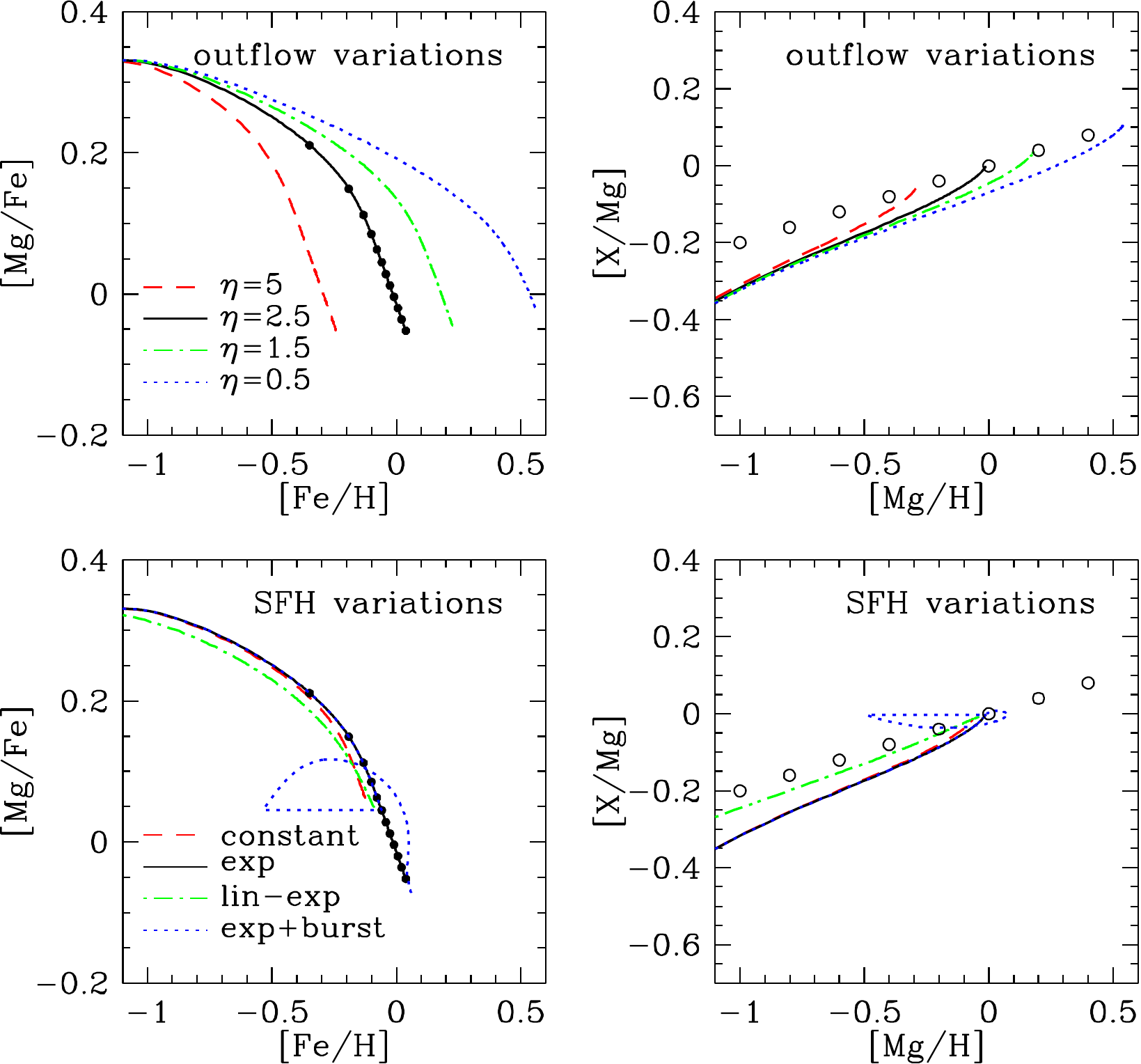}
\caption{Evolutionary tracks of $\mgfe$ vs. $\feh$ (left panels)
and $\xmg{X}$ vs. $\mgh$ (right panels) for a hypothetical CCSN element with a
metallicity-dependent yield (eq.~\ref{eqn:metdepyield}).
Upper panels show models with $\mdotstar(t) \propto e^{-t/\tausfh}$,
all with $\tausfh=6\Gyr$ but with four different values of the 
outflow mass-loading
parameter $\eta$ as marked.  Circles on the $\eta=2.5$ track in
the left panel mark $1\Gyr$ intervals ($t_{\rm max}=12\Gyr$).
Circles in the right panel show
the value of $\xmg{X}$ corresponding to the yield ratio 
$y_{\rm X}/y_{\rm Mg}$.  
Models in the lower panels have $\eta=2.5$ and four different
star formation histories: constant, exponential with $\tausfh=6\Gyr$,
linear-exponential ($\mdotstar(t) \propto t e^{-t/\tausfh}$) with
the same $\tausfh$, or exponential with a gas influx and subsequent
burst of star formation at $t=6\Gyr$.
The predicted trend of \xmg{X} with \mgh\ is only weakly dependent
on the outflow rate or star formation history, and it is close to
but steeper than the trend of $y_{\rm X}/y_{\rm Mg}$ with $\mgh$.
}
\label{fig:metdepyield1} 
\end{figure}

The top row of Figure~\ref{fig:metdepyield1} shows results for four
models, each of which assumes a constant star formation efficiency
timescale $\tau_* \equiv M_{\rm gas}/\mdotstar = 2\Gyr$ and
an exponential star formation history
$\mdotstar(t) \propto e^{-t/\tausfh}$ with $\tausfh=6\Gyr$.
Star formation drives an outflow with mass loading factor
$\eta = \dot{M}_{\rm out}/\mdotstar = 5$, 2.5, 1.5, or 0.5,
which ejects gas at the current ISM metallicity.
We adopt a $t^{-1.1}$ power-law delay time distribution for
SNIa \citep{Maoz2012,Maoz2012b} with a minimum delay time of $0.15\Gyr$.
The left panel shows evolutionary tracks in the $\mgfe-\feh$ plane.
As discussed by \cite{Andrews2017} and \cite{Weinberg2017}, the
abundances of one-zone models with constant parameters typically
approach late-time equilibrium values that depend primarily
on the element yield and the outflow parameter $\eta$.

The top right panel shows model tracks of $\xmg{X}$ vs. $\mgh$ for
an element with IMF-averaged yield
\begin{equation}
y_{\rm X} = y_{\rm Mg} \left(Z_{{\rm X},\odot}/Z_{{\rm Mg},\odot}\right)
            \times 10^{\alpha_{\rm cc}\mgh}~,
\label{eqn:metdepyield}
\end{equation}
where $Z_{{\rm X},\odot}$ is the solar abundance of element X and
we have adopted $\alpha_{\rm cc}=0.2$ for this example.
We assume that element X is produced entirely by CCSN, so
results in this panel are independent of the adopted iron yields
and SNIa delay time distribution.
At early times the ratio $Z_{\rm X}/Z_{\rm Mg}$ of newly forming
stars is lower than the yield ratio $y_{\rm X}/y_{\rm Mg}$, shown
by the circles, because 
much of the ISM enrichment came from stars with metallicity lower 
than the current ISM value.  However, once the $\mgh$ ratio 
approaches its equilibrium value, the $\xmg{X}$ track bends slightly
upward and settles at the value implied by the yield ratio.
The $\xmg{X}-\mgh$ tracks are therefore not pure power-laws, and
their effective slopes are steeper than those of the yield trend itself.
Models with exponential star formation history and
a yield slope of 0.20 produce a trend slope of 
$\alphacc \approx 0.35$, similar to that inferred for Mn
(see Figure~\ref{fig:twoprocess_peak}).
The model tracks end at different $\mgh$ because of the different
values of $\eta$, but where they overlap they are nearly identical.

The lower panels of Fig~\ref{fig:metdepyield1} show tracks for four
models that have $\eta=2.5$, $\taustar=2\Gyr$, and $\alpha_{\rm cc}=0.2$
but four different star formation histories.  The tracks for a 
constant SFR are nearly identical to those for an exponentially
declining SFR except that they terminate at a slightly lower
equilibrium metallicity.  We have experimented with shorter $\tausfh$
and different values of $\taustar$ and again find almost no
difference in the $\xmg{X}$ vs. $\mgh$ tracks except for changes in
endpoint.  The green dot-dashed curve shows a model with a
star formation history $\mdotstar(t) \propto t e^{-t/\tausfh}$
that rises linearly at early times and declines exponentially at 
late times, peaking at $t=\tausfh=6\Gyr$.  With a rising SFR,
the track of $\xmg{X}$ vs. $\mgh$ is closer to the yield ratio
because most of the ISM enrichment at a given time comes from
stars near the current ISM metallicity.  The blue dotted curve
shows a model that is identical to the fiducial exponential model
at early times but experiences an influx of pristine gas at $t=6\Gyr$,
which doubles the gas mass and lowers all abundances by a factor of two.
The doubled gas supply induces a burst of star formation that
temporarily enhances $\mgfe$ by boosting the CCSN rate relative
to the SNIa rate.  In the $\xmg{X}$ vs. $\mgh$ diagram, the primary
effect is an instantaneous excursion to lower $\mgh$ at fixed $\xmg{X}$,
with a return track that loops downward because the lower metallicity
reduces the yield $y_{\rm X}$.  Gas inflows and bursts of star formation
are thus a potential source of scatter in element ratios, but
even this fairly strong inflow/burst has a moderate impact
on the $\xmg{X}-\mgh$ track, and it will affect only those
stars formed during the burst.

\begin{figure}
  \includegraphics[width=0.40 \textwidth]{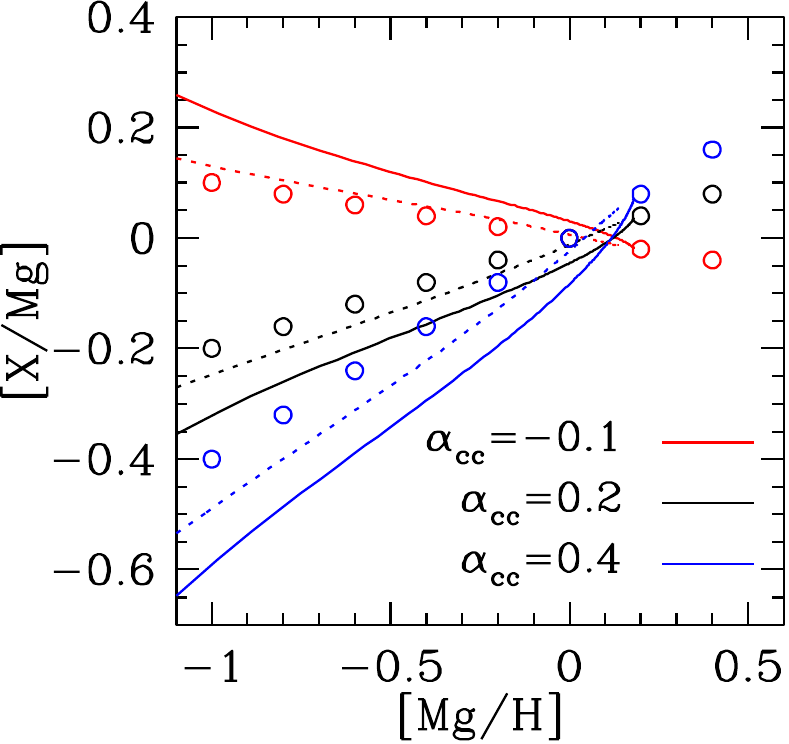}
\caption{Tracks of $\xmg{X}$ vs. $\mgh$ for three different values
  of the yield index $\alpha_{\rm cc}=-0.1$ (red), 0.2 (black), and 0.4 (blue).
Solid and dotted curves show models with exponential and linear-exponential
star formation histories, respectively, all with 
$\tausfh=6\Gyr$, $\tau_*=2\Gyr$, and $\eta=1.5$.
Circles show the values of $\xmg{X}$ corresponding to the yield ratio.
}
\label{fig:metdepyield2} 
\end{figure}

Figure~\ref{fig:metdepyield2} shows $\xmg{X}-\mgh$ tracks for
three different choices of the yield index and an exponential
or linear-exponential star formation history.  In each case
the track for a linear-exponential history is just slightly
steeper than the yield metallicity-dependence.  The track
for an exponential history is steeper by about 0.1, with 
effective slopes of approximately 0.55, 0.30, $-0.20$ for
$\alpha_{\rm cc} = 0.4$, 0.2, and $-0.1$.

From these numerical experiments we conclude (1) that the 
$\xmg{X}-\mgh$ track for a given metallicity-dependent yield
is only weakly sensitive to the chemical enrichment history, and
(2) that the $\xmg{X}-\mgh$ track is moderately steeper than
the actual metallicity dependence of the yield.
The first conclusion largely explains why median sequences
are nearly independent of Galactic position even for elements
with metallicity-dependent yields.
The second conclusion implies that the results of our
2-process model can be used as a first-cut empirical guide
to the metallicity dependence of nucleosynthetic yields for
the elements examined in this paper, though to test an
{\it ab initio} yield model it would be better to predict
median sequences directly and compare to our measurements.
We have also not examined the relation of $\alpha_{\rm Ia}$ to
the yield metallicity dependence for SNIa.
Our median sequences run from about $\mgh= -0.8$ 
to $\mgh = 0.45$, and the metallicity dependence in this regime cannot
necessarily be extrapolated to much lower metallicities, where
the stellar physics or nucleosynthesis physics could change
in important ways.

Our inferences apply to IMF-averaged yields.  
For some elements, the CCSN yield is expected to depend directly on metallicity
at a given supernova progenitor mass.
For essentially all elements, the expected CCSN 
yield at a given metallicity depends strongly on the 
supernova progenitor mass.\footnote{For illustrative 
plots of both points, see Appendix B of
\cite{Andrews2017}, based on the supernova yields of \cite{Chieffi2004}
and \cite{Limongi2006}.}
Even if the metallicity dependence at fixed progenitor mass is weak,
the IMF-averaged yield can depend on metallicity if the
IMF changes with metallicity or if the boundary between which
stars explode as supernovae and which collapse directly
to black holes \citep{Pejcha2015,Sukhbold2016} changes with 
metallicity.  The success of our model in explaining the observed
APOGEE sequences for even-$Z$ elements, with small inferred
values of $\alpha_{\rm cc}$, provides qualitative evidence against
changes in the IMF or in the mass ranges of exploding stars over
the lifetime and metallicity range of the disk.  However, specific
scenarios for such changes should be tested directly against the
observations.

\subsection{Offsets of median trends}
\label{sec:offsets}

\begin{figure}
\includegraphics[width=0.45 \textwidth]{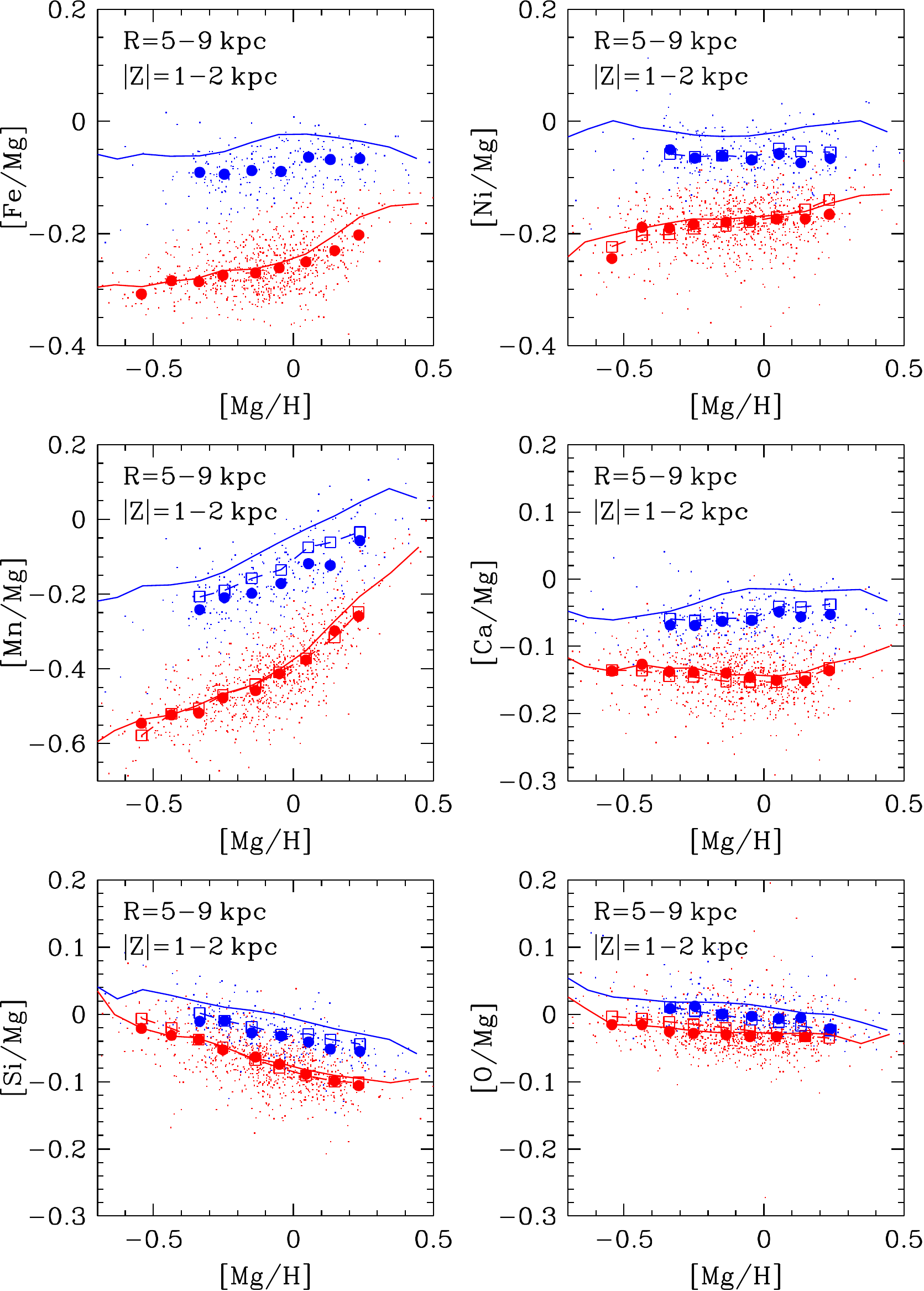} 
\caption{Offsets of median sequences at high $|Z|$ and interpretation
of these offsets in terms
of the 2-process model.  ({\it Top left}:) Light blue and red points
show \femg\ vs. \mgh\ for ${\rm SNR}>80$ 
stars at $R=5-9\kpc$ and $|Z|=1-2\kpc$ 
in the high-\femg\ and low-\femg\ populations, respectively.
Large filled circles show the median \femg\ in 0.1-dex bins of \mgh\ for bins
with at least 15 stars.  Solid lines show the median trends derived from
the full disk sample of ${\rm SNR}>200$ stars; at large $|Z|$, stars in
the high-\femg\ population have systematically lower \femg\ than stars
near the midplane, while the low-\femg\ sequence is nearly unchanged.
({\it Top right:}) Similar results for \xmg{Ni}.  Open squares and
connecting dashed lines show the median trend predicted by applying
the 2-process model to the median \femg\ values for the high-$|Z|$ stars.
({\it Other panels:}) Similar results for Mn, Ca, Si, and O.
Good agreement between the open squares and the filled circles
shows that the offsets of median \xmg{X} sequences at high $|Z|$, 
where they are seen, can be explained by the lower median
SNIa/CCSN ratio of these stars.
}
\label{fig:highZ} 
\end{figure}

As noted in \S\ref{sec:maps}, some elements show small offsets in
the median sequence for high-\femg\ stars far from the plane.
Figure~\ref{fig:highZ} examines this behavior more closely,
comparing the median sequences of ${\rm SNR}>80$ stars with
$R=5-9\kpc$ and $|Z|=1-2\kpc$ to the global median trends derived
from the full disk sample of ${\rm SNR}>200$ stars.
For \femg\ (top left panel), the median locus of the high-\femg\ population 
(blue points) is displaced downward by 0.03-0.07 dex, as seen 
previously in Figure~\ref{fig:FeMg}.
The low-\femg\ locus is only slightly shifted, with most points 
showing displacement less than 0.02 dex.  
The top right panel shows similar behavior for \xmg{Ni}.
Open squares show the predictions of the 2-process model,
with parameters inferred by fitting the full sample median sequences,
applied to the median \femg\ values of this high-$|Z|$ sample.
They fit the observed median sequences almost perfectly, 
implying that the slightly lower \xmg{Ni} values for high-$|Z|$
stars arise because these stars have a lower average SNIa/CCSN
ratio even within the high-\femg\ population.
The predicted (and observed) displacements are smaller than those 
for \femg\ because the SNIa contribution to Ni is smaller,
with $\Rxsun=0.59$ instead of 1.0.
Predicted (and observed) displacements are much smaller for
the low-\femg\ population because the $\femg-\mgh$ locus is
itself nearly unchanged.

The middle left panel shows similar results for \xmg{Mn}.  
Here the 2-process model explains most of the displacement of
the high-$|Z|$ median sequence, but it still slightly overpredicts
the \xmg{Mn} values.  The remaining panels show results for
three $\alpha$-elements: Ca, Si, and O.  The model fits imply
that Ca has the largest SNIa contribution, so it has the largest
predicted displacements of the high-$|Z|$ \xmg{X} locus, 
in good agreement with the data.  The model also successfully
predicts the smaller offset for \xmg{Si}.  Even the tiny
($0.01-0.02$ dex) drop in the \xmg{O} locus accords with the
small but non-zero $\Rxsun=0.07$ found for oxygen in the 
full sample fits.

\cite{Martig2016} find that {\it within} the high-\femg\ (low $\alpha/{\rm M}$)
population, higher $\alpha/{\rm M}$ ratios correlate with older stellar
ages (see their fig.~5).  It is plausible that these older stars have
a larger scale height because of the longer time available for dynamical
heating, so that they are preferentially represented in a high-$|Z|$
sample.  Relative to their mid-plane peers, the 
systematically smaller SNIa element fraction for these stars
lowers their \femg\ ratios, and our results show that it
also explains the offsets in median sequences at high $|Z|$ for other elements.
The smaller the SNIa contribution to a given element, the smaller
the offset.

The ability of the 2-process model to explain the systematic offsets
of high-$|Z|$ sequences is a reassuring sign that it is representing
the underlying physics that governs abundance ratios, not merely
fitting a data set with empirical parameters.  For well measured elements,
we also find that the dispersion of abundance ratios relative to
the star-by-star prediction of the 2-process model is smaller than
the dispersion relative to the median abundance ratio of stars
in the same population.  For example, for
${\rm SNR}>200$ stars, the rms scatter of \xmg{Ni} ratios relative 
to the median sequences is 0.038 dex, while the rms scatter
relative to the 2-process model predictions is 0.030 dex, a 37\%
reduction of variance.  In physical terms, the dispersion of
$\femg$ ratios at fixed \mgh\ within a given population represents
a dispersion of SNIa vs.\ CCSN enrichment, and other element ratios
respond accordingly.  Quantifying the intrinsic scatter of abundance
ratios relative to the 2-process model (or median trends) requires
careful assessment of the contribution of measurement errors to
the observed scatter; we defer this task to a future paper.

\section{Sequence fractions, MDFs, and metallicity gradients}
\label{sec:gradients}

Because the median trends we find are nearly independent of
location, our full multi-element cartography can be summarized
approximately by the combination of these global trends,
the relative contributions of low-\femg\ and high-\femg\ populations
as a function of Galactic position, and the distribution of \mgh\
for each population as a function of Galactic position.
In this section we turn our attention to the second and
third of these components.  While sample selection should have
minimal impact on the sequences of \xmg{X} at a given \mgh,
they can have a larger and more model-dependent effect in shaping
these popuation statistics.  Specifically, for a given star formation
history and enrichment history, the MDF of an evolved star sample
can differ from the MDF of long-lived main sequence stars because the
ratio of red giants to main sequence stars changes with the age
of the stellar population.
Modeling the inner Galaxy population of H15,
\cite{Freudenburg2017} find that the net impact on the MDF
is small, but in general one should forward-model the sample
selection to test a scenario of Galactic chemical enrichment.
Our statistics here should be taken as those of 
$1 < \log g < 2$ stars.  

\begin{figure*} 
   \includegraphics[width=\textwidth]{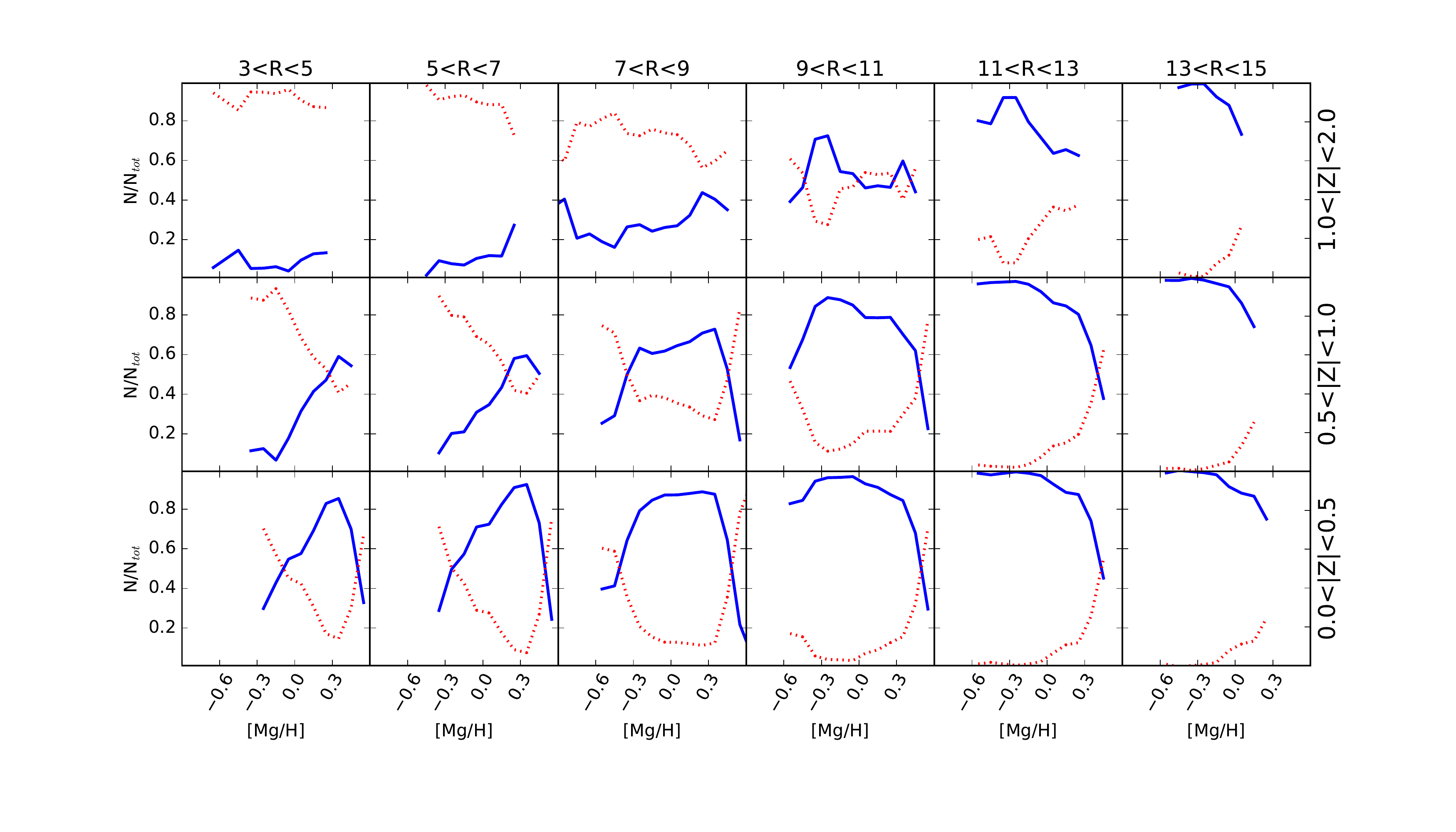}
\caption{Fraction of stars in the low-\femg\ population (red dashed) and
high-\femg\ population (blue solid) in 0.1-dex bins of \mgh\ in the
same Galactic zones used in Figs.~\ref{fig:FeMg}-\ref{fig:MnMg}.
Low-\femg\ (``chemical thick-disk'') stars are more prevalent at
high $|Z|$, small $R$, and low \mgh. Data are
shown only where the total number of stars per bin exceeds 30.
}
\label{fig:fracseq}
\end{figure*}

Figure~\ref{fig:fracseq} shows the fraction of low-\femg\ and high-\femg\
stars in bins of \mgh\ in each of our 18 Galactic zones.
As expected, the low-\femg\ (``chemical thick-disk'') population
dominates at high-$|Z|$, small $R$, and low-\mgh.  For $R > 5\kpc$
the high-\femg\ (``chemical thin-disk'') population dominates near
the midplane at nearly all \mgh.
In the outer Galaxy, $R > 11\kpc$, the ``chemical thin-disk'' population 
flares, and high-\femg\ stars dominate even 
at $|Z|=1-2\kpc$ \citep{Minchev2015,Minchev2017,Bovy2016,Mackereth2017}.
The high-\femg\ population usually dominates at super-solar \mgh,
but our division at $\mgfe=+0.12$ in this regime 
is somewhat arbitrary (see Figure~\ref{fig:MgFe}),
and the population fractions at $\mgh>0$ are sensitive to this choice.

\begin{figure*} 
   \includegraphics[width=\textwidth]{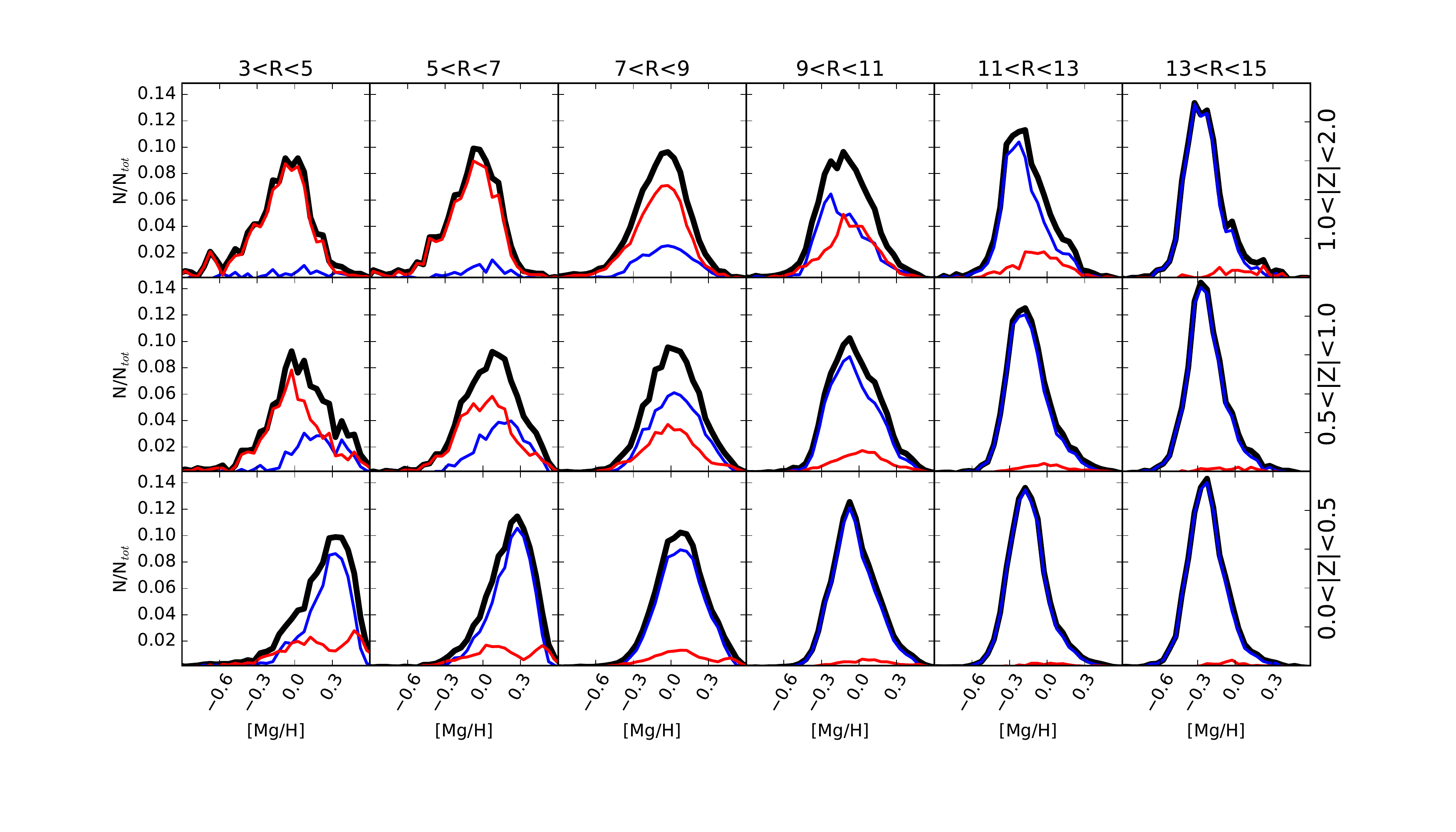}
\caption{Distributions of \mgh\ for all sample stars (black),
low-\femg\ stars (red) and high-\femg\ stars (blue) in the
same 18 Galactic zones shown in Figure~\ref{fig:fracseq}.
Distributions are computed in bins of 
0.1 dex in
\mgh, but they are plotted as curves rather than histograms for visual
clarity.  
Near the midplane, the full sample distribution 
shifts from negatively skewed at small $R$ to positively skewed
at large $R$, a trend that is present but weaker at high $|Z|$
(see Figure~\ref{fig:mdfstats} below).
}
\label{fig:mdf}
\end{figure*}

Figure~\ref{fig:mdf} shows the \mgh\ MDFs in each of our 18 zones, for
the full sample and for the low-\femg\ and high-\femg\ populations
individually.  Our results are similar to those of H15, who showed 
distributions of \feh\ and [$\alpha$/H] in the same zones.
Most strikingly, the midplane MDF changes shape from negatively
skewed in the inner Galaxy to approximately symmetric at the
solar annulus to positively skewed in the outer Galaxy.
As discussed by H15, the inner Galaxy form resembles the prediction
of generic one-zone evolution models, and the symmetric or
positively skewed forms in the outer Galaxy can arise if radial
migration from smaller $R$ is responsible for populating the 
high-metallicity tails.  The MDF at $|Z|=1-2\kpc$ is roughly
symmetric for $R<9\kpc$, though it too is positively skewed
at larger $R$.  For the low-\femg\ population, the change of shape
is much less pronounced than for the high-\femg\ population.

\begin{figure}
\includegraphics[width=0.45 \textwidth]{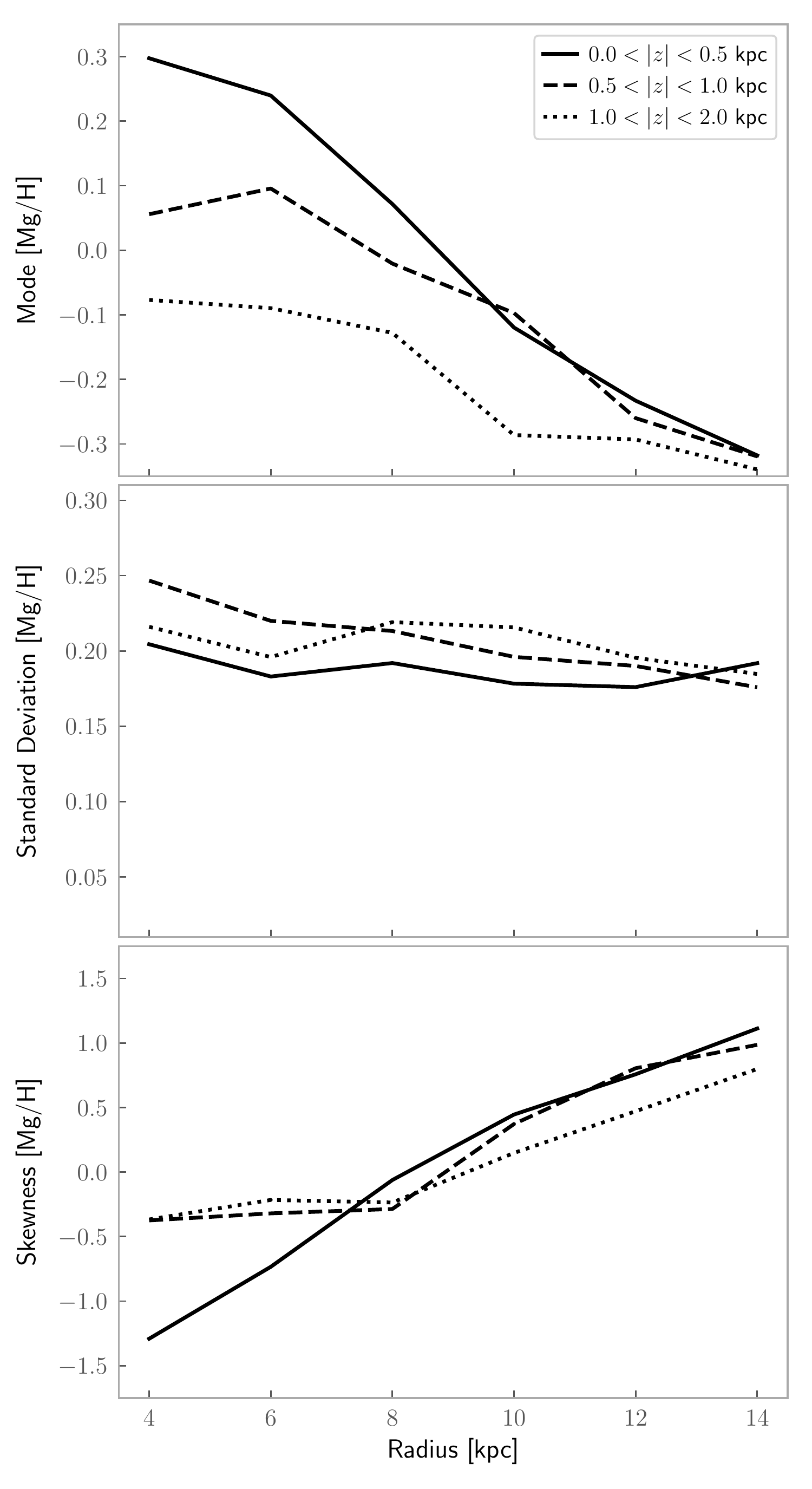}
\caption{Mode (top panel), dispersion (middle panel), and 
skewness (bottom panel) of the \mgh\ MDF as a function of 
Galactocentric radius, for the all-star distributions
(black histograms) of Figure~\ref{fig:mdf}.  
Solid, dashed, and dotted lines show results for
$|Z|/\kpc = 0-0.5$, $0.5-1$, and $1-2$, respectively.
}
\label{fig:mdfstats}
\end{figure}

Figure~\ref{fig:mdfstats} summarizes and quantifies these trends
by plotting the mode, dispersion, and skewness of the \mgh\ 
distribution as a function of radius, for all ${\rm SNR}>80$ stars
with $\mgh > -0.8$ in each of the three $|Z|$ bins.  
To reduce noise in the estimates, the mode is computed after applying
kernel density estimation to create a smoothed histogram from a fine-grained
\mgh\ distribution; the dispersion and skewness are computed directly
from the central moments of the stellar distribution with no binning
or smoothing.
Near the midplane, the modal metallicity declines from 
$\mgh=+0.3$ at $R=4\kpc$ to $\mgh=-0.3$ at $R=14\kpc$,
exhibiting a roughly constant gradient of $-0.06\,{\rm dex}\kpc^{-1}$.
Gradients are shallower at large $|Z|$ because the modal \mgh\ 
is smaller in the inner Galaxy.  
The dispersion of the MDF is approximately
constant with radius, with $\sigma_{\mgh} \approx 0.2-0.25\,$dex for
$|Z| < 0.5\kpc$ and a slightly higher $\sigma_{\mgh} \approx 0.25-0.3\,$dex
for $|Z|=1-2\kpc$.  The skewness plot shows the reversal of the
midplane MDF shape between $R=6\kpc$ and $R=12\kpc$, with
skewness changing from $-1.3$ to $+1.0$.  
At $|Z|=0.5-2\kpc$ the skewness is roughly flat at $\approx -0.4$ out
to $R=8\kpc$, then rises steadily to $\approx +0.7$ at $R=14\kpc$.
The values of the skewness are sensitive to the adopted lower
cutoff at $\mgh=-0.8$; including lower metallicity stars reduces
the skewness values, though it does not change the qualitative
trends.

Because the MDF shape changes with radius, the derived metallicity
gradient depends on what statistic one uses to characterize the
stellar metallicity.  The mode is arguably the best choice, 
because it likely corresponds to the typical metallicity of stars
formed {\it in situ} at each radius, while the tails of the MDF,
and thus the mean of the distribution, will be more affected by
radial migration of stars.  Our mode gradient of $-0.06\,{\rm dex}\kpc^{-1}$
is comparable to that derived from Cepheid stars, approximately 
$-0.045\,{\rm dex}\kpc^{-1}$ for \mgh\ \citep{Genovali2015} and
$-0.05\,{\rm dex}\kpc^{-1}$ for \feh\ \citep{Luck2018}.
Gas phase oxygen abundances of HII regions yield similar
estimates of radial gradients, e.g., 
$-0.06\,{\rm dex}\kpc^{-1}$ from \citep{Rudolph2006}.
In principle, the median or mean metallicity gradient of the
full stellar population could differ significantly from that of
gas and young stars, but the metallicity gradients found in the
Gaia-ESO survey by \cite{Mikolaitis2014} are at most slightly shallower,
$-0.045\,{\rm dex}\kpc^{-1}$ for [M/H].

\section{Conclusions and Prospects}
\label{sec:conclusions}

Extending earlier work on ``chemical cartography'' of the Galactic disk
with APOGEE (\citealt{Hayden2014,Nidever2014}; H15), we have examined
the distribution of stellar abundance ratios as a function of 
Galactocentric radius $R$ and midplane distance $|Z|$, using
20,485 evolved stars selected from the SDSS DR14 APOGEE 
data set.  For the elements considered in this paper, we find that the
description of multi-element cartography in the Milky Way disk is
approximately separable.  The proportion of stars on the low-$\femg$ 
(``high-$\alpha$'') and high-$\femg$ (``low-$\alpha$'') sequences
varies strongly with $R$, $|Z|$, and $\mgh$.  However, along each sequence
the median $\xmg{X}$ ratio depends only on $\mgh$, with at most
subtle dependence on Galactic position.  The position-dependent
distribution in the $\femg-\mgh$ plane constrains
the history of chemical enrichment and stellar radial migration
across the disk.  The nearly universal $\xmg{X}-\mgh$ trends
provide insights on nucleosynthesis by core collapse
and Type Ia supernovae.

\subsection{Mg as a reference element}

This separability of multi-element cartography is not evident if 
one simply plots $\xfe{X}-\feh$ distributions in different 
Galactic zones.  The relation appears in our analysis because we adopt a 
pure CCSN element, Mg, as our reference, and because we separately
examine trends in the low-\femg\ and high-\femg\ populations.
These choices suppress misleading trends that would otherwise
arise because of the changing SNIa contribution to \feh.
We would obtain similar results if we used O, or a weighted
mix of $\alpha$-elements, in place of Mg, but we have chosen
Mg as our reference to allow simpler examination of trends 
within the $\alpha$-elements and straightforward comparison to
other surveys (which usually measure Mg more accurately than O).
Standard nucleosynthesis calculations predict that Mg is
produced almost entirely by CCSN and that its IMF-averaged
yield is nearly independent of metallicity \citep{Andrews2017}.

\subsection{Position-dependent {\rm [Fe/Mg]-[Mg/H]} distributions}

For the \femg-\mgh\ distributions (summarized in
Figs.~\ref{fig:fracseq}-\ref{fig:mdfstats}), our results
are qualitatively similar to those of H15, who used the
DR12 APOGEE data set and the \am\ and \mh\ ratios.
Like H15 and \cite{Nidever2014} we find that the locus traced
by low-\femg\ (high-\am) stars is nearly constant in all
Galactic zones where we have enough stars to measure it.
The fraction of low-\femg\ stars increases with $|Z|$ at
fixed $R$ --- the well-known ``chemical thick-disk''
\citep{Bensby2003}.  At fixed $|Z|$, the low-\femg\ 
fraction is highest in the inner Galaxy and decreases outward.
For the inner Galaxy, the mode of the MDF shifts to higher
metallicity when moving from high $|Z|$ to the midplane,
a result interpreted by \cite{Freudenburg2017} as a sign
of ``upside down'' disk formation \citep{Bird2013}.  
Near the midplane, the mode of the MDF moves to lower \mgh\ 
with increasing $R$, the well known radial metallicity gradient.
The gradient is progressively weaker, though still present,
at larger $|Z|$, as predicted in chemical evolution models where
the high-$|Z|$ population is comprised of older stars
(\citealt{Minchev2014}, see their fig.~10).
The rms width of the MDF is roughly independent
of position, but the shape changes from negatively skewed at small $R$ to
positively skewed at large $R$, a trend that is most obvious
near the midplane.  H15 interpret this change of shape as a sign
of stellar radial migration --- in particular, the high metallicity
tail of the MDF at large $R$ is comprised of stars that formed 
in the inner Galaxy (see, e.g., \citealt{Minchev2013}, fig.~3).
These results provide rich and informative tests for Galactic
chemical evolution models and hydrodynamic cosmological simulations.
Recent simulation studies suggest that explaining the Milky Way's
observed bimodal \afe-\feh\ distributions is a significant
challenge \citep{Grand2018,Mackereth2018}.

\subsection{Implications for nucleosynthetic yields}

The approximate constancy of the median $\xmg{X}-\mgh$ sequences
allows us to combine the full disk 
red giant sample and select only those stars with ${\rm SNR} > 200$
spectra, yielding the distributions shown in 
Figs.~\ref{fig:alpha_2alpha}-\ref{fig:fepeak_2alpha} and
the median sequences tabulated in 
Tables~\ref{tbl:medseq_alpha_lowFe}-\ref{tbl:medseq_peak_highFe}.
To interpret these sequences we have developed a 
semi-empirical ``2-process'' nucleosynthesis model that assumes
(a) Mg is produced purely by CCSN with a metallicity-independent
IMF-averaged yield, 
(b) Fe is produced by CCSN and SNIa with a metallicity-independent yield,
(c) other elements are the sum of a CCSN process and an SNIa process,
each of which may have a power-law dependence of (X/Mg) on (Mg/H),
with a relative amplitude $\Rxsun$ at solar \mgh.
The fits of this model to the observed median
sequences are shown in Figs.~\ref{fig:twoprocess_alpha}-\ref{fig:twoprocess_peak}.
As shown in Figs.~\ref{fig:metdepyield1}-\ref{fig:metdepyield2}, the
derived power-law slope $\alphacc$ 
can be interpreted approximately as the implied
metallicity dependence of the IMF-averaged CCSN yield in the sample
range $-0.8 \leq \mgh \leq +0.4$, though the observed slope is 
somewhat steeper than the true yield slope by an amount that depends,
slightly, on the star formation history.  Constraints on the SNIa
slope $\alphaIa$ are
often weak, and they are sensitive to the assumption of a power-law
form for the CCSN dependence, so predictions of metallicity dependent
SNIa yields should be tested by constructing a full forward model and
allowing for uncertainties in star formation history and in 
the CCSN yield model.
The measurement of the 
relative amplitude $\Rxsun$ is more robust, as it follows directly
from the separation of the low-\femg\ and high-\femg\ sequences
in the $\xmg{X}-\mgh$ diagram.  

Our median sequences and 2-process fits provide a trove of quantitative
constraints for supernova nucleosynthesis models.  Many of the results
are as expected.  Among the $\alpha$-elements, O shows little SNIa
contribution, while Si and Ca show progressively larger SNIa components.
Among light odd-$Z$ elements, Al and K are dominated by CCSN, with a
positive metallicity-dependent yield.  The iron-peak elements show 
large SNIa contributions, with minimal metallicity dependence for
the even-$Z$ elements (Cr, Fe, Ni) and strong metallicity dependence
for the odd-$Z$ elements (V, Mn, Co).  Of the elements considered
in this paper, Mn has the largest SNIa contribution and the strongest
metallicity dependence of the median sequences.

Some of our results are more surprising.  The two median sequences
for \xmg{S} are nearly identical, implying almost no SNIa contribution
to sulfur abundances.  The implied metallicity dependence of Si yields
is perhaps unexpectedly strong, while that of K is perhaps unexpectedly 
weak.  
Both \xmg{Na} and \xmg{P} show large separations of the
low-\femg\ and high-\femg\ sequences, implying large SNIa contributions
to these elements.  Our inferred SNIa contribution to Ni is substantially
lower than for Fe, with $R_{\rm Ia}^{\rm Ni}=0.6$ vs.
$R_{\rm Ia}^{\rm Fe}=1.0$.

\subsection{Extensions and applications of the 2-process model}

Our model does not allow for contributions from other nucleosynthetic
sources, such as AGB stars.  To the extent that the time delay distribution
for AGB enrichment mimics that of SNIa enrichment, an AGB contribution
could masquerade as an SNIa contribution in our model.  However,
the AGB contribution to the elements considered here is expected
to be small.  The small offset in \xmg{O} between the
low-\femg\ and high-\femg\ sequences could plausibly 
represent an AGB contribution to oxygen, though it could also arise
from model-dependent systematics in the APOGEE oxygen measurements.
We have not considered more exotic scenarios in which, e.g., the
stellar IMF or the mass range of CCSN progenitors or the properties of
SNIa explosions change substantially with time or metallicity.
The success of our simple model provides some generic evidence against
such scenarios --- for example, a change in the high-mass IMF would
be expected to change the ratios of explosive nucleosynthesis
elements (Si, S, Ca) to the hydrostatic elements O and Mg
\citep{McWilliam2013,Vincenzo2015,Carlin2018}.
However, any specific model must be tested for its ability
to reproduce the trends observed here.

The 2-process model can be applied to individual stars as well as 
median sequences, predicting a star's \xmg{X} ratios given only its
\mgh\ and \femg\ measurements.  Specifically, one uses
Equation~(\ref{eqn:aratio}) to infer the star's relative SNIa
enrichment contribution $\AIa/\Acc$, then uses 
Equation~(\ref{eqn:xmg}) to predict \xmg{X};
the predicted value of \xh{X} can be obtained by adding
the measured \mgh.  In future work we will examine
both the scatter in \femg\ at fixed \mgh\ (see \citealt{Bertran2016}
for a recent study of \xfe{O} scatter) and correlated star-by-star
deviations from the predictions of the 2-process model.  The former
provides constraints on the stochasticity of star formation in the
galaxy, since bursts or oscillations in the star formation rate
drive variations in the ratio of CCSN to SNIa enrichment
(\citealt{Weinberg2017}; J.W. Johnson et al., in prep.).
The latter could reveal signatures of additional nucleosynthesis
channels beyond the IMF-averaged CCSN and SNIa processes in the
current model.  By removing two known strong effects on elemental
abundance patterns --- the ratio of CCSN to SNIa enrichment and the
metallicity-dependent yields of some elements --- the 
deviation-from-model approach may allow one to tease information 
from subtler observational clues.  

In the longer term we aim to apply
these methods to a wider range of elemental abundances and thus probe
a larger set of nucleosynthesis channels.  This can include
APOGEE C and N abundances for main sequence or sub-giant stars that are 
not affected by mixing on the giant branch, the $s$-process
elements Nd and Ce that are detectable in APOGEE spectra
\citep{Hasselquist2016,Cunha2017}, and the many elements
accessible in optical spectroscopic samples.
We also plan to investigate whether
bulge and halo star samples show the same supernova nucleosynthesis
patterns found here for disk stars, or instead exhibit differences
that could be a sign of changing IMFs or radically different 
yields at low metallicity.

A basic challenge of chemical evolution modeling is that abundance
patterns depend on both the enrichment history of the host galaxy
and on nucleosynthetic yields that are difficult to predict robustly
from first principles.  The approximate separability found here suggests
one way to make progress, by analyzing abundance distributions in a way
that isolates empirical constraints on nucleosynthesis from the 
broad-brush effects of galactic-scale evolution.   To the extent that
we can decompose star-by-star abundance patterns into sums of individual
processes, the amplitudes of these processes may themselves be the
best quantities to use in future chemical cartography, 
charting the contributions
of distinct enrichment channels in different zones of the Galaxy.
Precise kinematic information from {\it Gaia} also makes it feasible
to change mapping variables from present-day $R$ and $|Z|$ to 
more intrinsically informative quantities such as guiding center
radius, orbital ellipticity, and vertical action.
The continuing explosive growth of Galaxy-scale 
multi-element abundance surveys offers bright prospects for
uncovering the physical origin of the elements and the history
of the Milky Way.

\acknowledgments

We thank Brett Andrews, Jan Rybizki, and Tuguldur Sukhbold
for valuable discussions of supernova yields and comments
on the draft manuscript.  Work by DHW, JAH, and JAJ was supported 
in part by NSF grants AST-1211853 and AST-1109178.
DAGH and OZ acknowledge support provided by the Spanish Ministry of
Economy and Competitiveness (MINECO) under grant AYA-2017-88254-P.
HJ acknowledges support from the Crafoord Foundation, Stiftelsen Olle 
Engkvist Byggm\"astare, and Ruth och Nils-Erik Stenb\"acks stiftelse.
SzM has been supported by the Premium Postdoctoral
Research Program of the Hungarian Academy of Sciences, and by the Hungarian
NKFI Grants K-119517 of the Hungarian National Research, Development and
Innovation Office.

Funding for the Sloan Digital Sky Survey IV has been provided by the
Alfred P. Sloan Foundation, the U.S. Department of Energy Office of
Science, and the Participating Institutions. SDSS acknowledges support
and resources from the Center for High-Performance Computing at the
University of Utah. The SDSS web site is www.sdss.org.

SDSS is managed by the Astrophysical Research Consortium for the
Participating Institutions of the SDSS Collaboration including the
Brazilian Participation Group, the Carnegie Institution for Science,
Carnegie Mellon University, the Chilean Participation Group, the French
Participation Group, Harvard-Smithsonian Center for Astrophysics,
Instituto de Astrof\'isica de Canarias, The Johns Hopkins University,
Kavli Institute for the Physics and Mathematics of the Universe
(IPMU) / University of Tokyo, Lawrence Berkeley National Laboratory,
Leibniz Institut f\"ur Astrophysik Potsdam (AIP), Max-Planck-Institut
f\"ur Astronomie (MPIA Heidelberg), Max-Planck-Institut f\"ur Astrophysik
(MPA Garching), Max-Planck-Institut f\"ur Extraterrestrische Physik (MPE),
National Astronomical Observatories of China, New Mexico State University,
New York University, University of Notre Dame, Observat\'orio Nacional /
MCTI, The Ohio State University, Pennsylvania State University, Shanghai
Astronomical Observatory, United Kingdom Participation Group, Universidad
Nacional Aut\'onoma de M\'exico, University of Arizona, University of
Colorado Boulder, University of Oxford, University of Portsmouth,
University of Utah, University of Virginia, University of Washington,
University of Wisconsin, Vanderbilt University, and Yale University.

\bibliographystyle{apj}
\bibliography{ref}

\end{document}